\documentclass[10pt,a4paper]{article}
\usepackage[utf8]{inputenc}
\usepackage{abstract} 
\usepackage[ngerman,english]{babel}
\usepackage[T1]{fontenc}
\usepackage{multirow}
\usepackage[table]{xcolor}
\usepackage{tabularx}
\usepackage{amsmath}
\usepackage{amsfonts}
\usepackage{amssymb}
\usepackage{braket}
\usepackage{cite}
\usepackage{enumitem}
\usepackage{framed}
\usepackage{setspace}
\usepackage{todonotes}
\usepackage{wrapfig}
\usepackage{rotating}
\usepackage{subcaption}
\usepackage{tcolorbox}
\usepackage{makeidx}
\usepackage{pgfplots}
\usepackage{pdflscape}
\usepackage{epstopdf}
\usepackage{color}
\usepackage{titling} 
\usepackage{blindtext} 
\usepackage{soul}
\usepackage{todonotes}
\usepackage[font=footnotesize]{caption}
\usepackage{graphicx}
\usepackage{anysize}
\usepackage[pdfborder={0 0 0}]{hyperref}
\usepackage[T1]{fontenc}
\usepackage{fancyhdr}

\marginsize{3.5cm}{2.5cm}{1cm}{1cm}

\newlength\figureheight 
	\newlength\figurewidth

\pretitle{\begin{center}\Large\bfseries} 
\posttitle{\end{center}} 
\title{Forward model for quantitative pulse-echo speed-of-sound imaging} 
\author{%
\textsc{Patrick Stähli, Maju Kuriakose, Martin Frenz*, Michael Jaeger} \\[1ex] 
\normalsize Institute of Applied Physics, University of Bern, Sidlerstrasse 5, 3012 Bern, Switzerland \\ 
\normalsize June 24, 2019\\
\normalsize *Corresponding author: martin.frenz@iap.unibe.ch
}
\date{} 
\renewcommand{\maketitlehookd}{%
\begin{abstract}
\noindent Computed ultrasound tomography in echo mode (CUTE) allows determining the spatial distribution of speed-of-sound (SoS) inside tissue using handheld pulse-echo ultrasound (US). This technique is based on measuring the changing phase of beamformed echoes obtained under varying transmit (Tx) and/or receive (Rx) steering angles. The SoS is reconstructed by inverting a forward model describing how the spatial distribution of SoS is related to the spatial distribution of the echo phase shift. CUTE holds promise as a novel diagnostic modality that complements conventional US in a single, real-time handheld system. 
Here we demonstrate that, in order to obtain robust quantitative results, the forward model must contain two features that were not taken into account so far: a) the phase shift must be detected between pairs of Tx and Rx angles that are centred around a set of common mid-angles, and b) it must account for an additional phase shift induced by the error of the reconstructed position of echoes. In a phantom study mimicking liver imaging, this new model leads to a substantially improved quantitative SoS reconstruction compared to the model that has been used so far. The importance of the new model as a prerequisite for an accurate diagnosis is corroborated in preliminary volunteer results.\\ 
\\
\\
\textbf{Keywords:} ultrasound tomography, reflection mode, multimodal imaging, inverse problem
\end{abstract}
}

\begin{document}

\maketitle



\section{Introduction}
\label{Sec:Introduction}
Ultrasound (US) is an integral part of today's clinical diagnostic practice since it provides us with real time display and flexible free-hand probe guidance while using non-ionizing radiation. The compact, portable and comparably inexpensive US systems make their use favourable  for general practitioners, emergency units and bed-side care. On the downside, classical gray-scale B-mode US often suffers from low sensitivity and non-specific contrast, resulting in difficulties in differential diagnosis for certain disease types \cite{baker1999sonography, konno2001liver, rahbar1999benign}. To complement B-mode images with additional structural and functional information in a single multi-modal system, much effort has been placed in recent years in developing new ultrasound-based modalities. This includes Doppler flow imaging, ultrasound elastography \cite{athanasiou2010breast,bamber2013efsumb,barr2010real,cosgrove2013efsumb,dietrich2017efsumb, sigrist2017ultrasound} and optoacoustic imaging \cite{hu2010photoacoustic, jaeger2012deformation, held2016multiple, Ulrich2018SpectralCorrection}.\\
Based on the dependence of the SoS on the tissues mechanical properties, SoS imaging is another promising modality that can help identifying disease-related changes of tissue composition and structure.  




Breast ultrasound computed tomography (UCT) showcases the potential of SoS imaging on the example of breast cancer diagnosis \cite{ruiter2013first}. In UCT, US is through-transmitted through the breast from many angles, and the spatial distribution of SoS is reconstructed based on analysing the detected US signals. In the straight ray approximation of ultrasound propagation, the time-of-flight (ToF) of pulsed signals is assigned to line integrals of slowness (inverse of SoS) along straight lines connecting sender and receiver locations, linking the ToF data to a Radon transform of the slowness distribution \cite{greenleaf1981clinical, jago1991experimental, zografos2013novel, carson1981breast}. This allows a fast linear reconstruction of the SoS, e.g. via the filtered backprojection, but with the disadvantage of a low spatial resolution due to diffraction and refraction. Refraction can be accounted for in an iterative way in a bent-ray ToF approach. Diffraction, however, requires methods that use the full signal as opposed to the ToF. Diffraction tomography based on the 1st order Born approximation is linear and fast, but limited to low contrast SoS variations so that its application to the breast requires a SoS prior with "reasonable low resolution" \cite{huthwaite2011high}. The best in-vivo images were so far obtained using non-linear iterative full-wave inversion schemes, but with the disadvantage of a high numerical cost \cite{wiskin2012non, sandhu2015frequency}.\\
Whereas UCT is able to achieve high resolution and quantitative SoS images of breast cancer, the big disadvantage is that it is a through-transmission technique, which limits its use to acoustically transparent parts of the human body. To provide SoS imaging with all the flexibility of conventional handheld US and without the need for specialized equipment, SoS imaging must be based on echo US for a one-sided access. This would allow to not only image the SoS in transparent parts of the body, but also in any other part that is routinely examined using echo US, e.g. for the diagnosis of cancer other than in the breast, of fatty liver diseases or for the assessment of plaque composition inside the carotid artery.\\
Reflection-mode SoS imaging is feasible when taking into account a multiple-scattering process beyond the 1st order Born approximation. This problem can in principle be solved as in UCT in a non-linear full-wave inversion. Even though promising theoretical results were obtained using this approach in 2D digital breast phantoms \cite{hesse2013nonlinear}, no clinical results have been presented so far, potentially because it requires low frequencies outside the bandwidth of conventional clinical US probes.  
Various alternative approaches were investigated that reconstruct the SoS based on pulse-echo signals. Techniques that estimate the average SoS between the transducer and the focal depth reported accurate SoS measurements of uniform tissue \cite{shin2010estimation,krucker2004sound,imbault2017robust}. These methods, however, have low accuracy in the presence of SoS inhomogeneities. The crossed-beam tracking method is based on measuring the round-trip time through the intersection of two scanning beams (one for transmit, one for receive), and provides a spatial resolution on the order of 10 mm  \cite{kondo1990evaluation,cespedes1992feasibility}. Recently, an approach was proposed that reconstructs the local SoS based on axial variations of the average SoS determined by optimizing the transmit and receive focusing. The technique was verified in phantoms without lateral SoS variations, where it provided an axial resolution of about 7 mm \cite{jakovljevic2018local}. 

We recently developed a reflection-mode technique, named computed ultrasound tomography in echo-mode (CUTE), which allows a real time determination of the SoS with promising spatial and contrast resolution \cite{jaeger2015computed, jaeger2015towards}. CUTE is based on analysing the spatial distribution of the echo phase in beamformed (using e.g. conventional delay-and-sum algorithm) radio-frequency (rf) mode US images. A deviation of the true SoS from the value assumed for beamforming results in a mismatch between the anticipated and the actual round-trip time of US propagation (henceforth termed ‘aberration delay’). A changing value of the aberration delay when detecting echoes under varying angles of ultrasound transmission and/or reception consequently results in a phase shift of these echoes, which is quantified in a spatially resolved way by e.g. Loupas type phase correlation\cite{loupas1995axial}. This concept is closely related to approaches that have been developed for aberration correction of superficial SoS variations that act like a phase-distorting screen directly in front of the transducer array. They are based on analysing the differential echo phase as function of transducer element position in the channel data \cite{flax1988phase, nock1989phase, rachlin1990direct, li1997phaseA, li1997phaseB, haun2004overdetermined}. CUTE goes beyond these approaches: By determining the phase shift in the beamformed images as opposed to the channel data, lateral resolution of phase shift data is achieved also away from the aperture. Based on a model of how the spatial distribution of SoS relates to the spatially resolved phase shift, the former can thus be reconstructed with a spatial resolution on the order of a couple of mm. Apart from serving as a source of diagnostic information, knowledge of the spatial distribution of SoS allows for aberration correction beyond the phase screen assumption \cite{jaeger2015full}. 

The core of CUTE is the forward model that relates the SoS to echo phase shift, and this forward model is closely linked to the data acquisition scheme. Previously, we proposed an implementation of CUTE where the rf-mode images were acquired under a variety of transmit (Tx) angles, whereas the echoes were detected with a constant receive (Rx) aperture \cite{jaeger2015computed}. The model (henceforth termed 'old model') was based on following key assumptions: 

\begin{enumerate}
\item Because the Rx aperture is constant, the echo phase shift is entirely determined by the changing aberration delay along the changing Tx propagation directions.
\item The echo phase shift is proportional to the difference in aberration delay along different round-trip paths.
\end{enumerate}

In \cite{jaeger2015computed,jaeger2015towards,jaeger2015full}, this forward model was formulated in the frequency-domain (FD). The FD formulation, however, cannot account for the absence of phase shift data in regions of low echo intensity, which leads to artefacts in the SoS image. Therefore we proposed that quantitative imaging requires a space-domain (SD) approach \cite{MJTalk2015}. Sanabria et al. verified in a simulation study the advantage of using a SD instead of a FD approach \cite{sanabria2018spatial}. However, in spite of promising results when using the SD approach, we realized that the old model leads to inconsistent results, depending on the sample geometry. \\

Here we present a new, more general model that solves this shortcoming by modifying the two key assumptions of the old model as follows:
\begin{enumerate}
\item Even if the Rx aperture is constant while changing the Tx angle, the echo phase shift contains the influence of a virtually changing Rx angle. To avoid ambiguities, we propose to switch from a pure Tx-steering approach to simultaneously steering both, the Tx- and the Rx- angles, around a variety of common mid-angles, in an approach similar to the common mid-point gather that has been used in seismic imaging \cite{bednar2009modeling} as well as in US \cite{li1997phaseA, li1997phaseB, haun2004overdetermined}. 
\item The \textit{a priori} unknown SoS distribution leads to an error of the reconstructed position of echoes. This error not only depends on the aberration delay, but independently also on the values of the steering angles. As a result, the initially assumed simple proportionality between the aberration delay and phase shift is no longer valid.
\end{enumerate}

In the theory section, we present a theoretical revision of the old model and the development of the new model. In a phantom study, we then compare the old and the new model in layered structures mimicking the abdominal wall and liver in preparation of one of our target clinical applications, the diagnosis of liver disease. The presented results reveal that the new model results in substantially improved quantitative SoS imaging among different layered phantom geometries. The article concludes with a discussion of the implications of the new model for the clinical use of CUTE, corroborated by a volunteer result that demonstrates the promise of the new model for a quantitative diagnosis of liver SoS. 


\section{Theory}
The theory was developed assuming a linear array probe and a 2D geometry of SoS distribution and sound propagation, but can be readily adapted to curved arrays for 2D sector scans, as well as to matrix arrays for 3D imaging. Fig. \ref{Fig:Theory1}(a) illustrates the measurement geometry. A linear US probe (parallel to coordinate x) is placed on top of the tissue (at coordinate z = 0), and transmits ultrasound pulses using a variety of different transmit settings, indexed by $n$. This can either be a variety of different Tx angles as proposed in earlier studies \cite{jaeger2015computed, jaeger2015towards}, or a sequential activation of elements or groups of elements. For the sake of visual clarity, a sequential activation of single elements is considered in Fig. \ref{Fig:Theory1}(a). Two different elements (indexed by $j_n$ and $j_{n^\prime{}}$) transmit an US pulse, one at a time. The two divergent wave fronts reach a point $\textbf{r}^\prime=(x^\prime{},z^\prime{})$ inside the tissue from two different angles $\phi_n$ and $\phi_{n^\prime{}}$. Intrinsic ultrasound reflectors located in a vicinity around $\textbf{r}^\prime{}$ lead to echoes that propagate back to the surface where they are detected by the different sensor elements (index $k$) of the array probe resulting in radio frequency (rf) signals $s(t,n,k)$ with the time indexed by $t$.
\\
In line with previous studies \cite{jaeger2015computed,jaeger2015towards,jaeger2015full}, we propose that a separate rf-mode image $u(\textbf{r},n)$ is reconstructed from the detected signals for each Tx setting $n$. The images are reconstructed from the complex (analytic) rf-signal (crf-signal), where the imaginary part of the amplitude is the Hilbert transform (along $t$) of the initial real-valued rf-signal. The complex signal generated by a point reflector located at $\textbf{r}^\prime$ can be modelled as a product of a complex exponential carrier modulated with a complex-valued envelope $G$: 

\begin{equation}
s(t,n,k)=G(t-t_0(\textbf{r}^\prime,n,k),\textbf{r}^\prime,k)\cdot exp \left[ 2\pi i f_0 \left (t-t_0(\textbf{r}^\prime,n,k) \right) \right]
\label{Equ:Theory_ComplexSignalGenerated}
\end{equation}
where $f_0$ is the centre frequency, and $t_0$ is the actual round-trip time of the echo. The function $G$ characterises the spatio-temporal impulse response of the system, and depends on $\textbf{r}^\prime$ as well as on the receiving element $k$. The dependence on $k$ takes into account that the signal may be influenced by the angle-dependent receive response of the elements, but also by spatial variations of SoS. The reconstructed crf-amplitude in a point $\textbf{r}=\textbf{r}^\prime$ when using delay-and-sum beamforming (DAS) is:

\begin{equation}
\begin{split}
u(\textbf{r}=\textbf{r}^\prime,n) &	= \sum_k \left[ s\left( \hat{t}_0(\textbf{r},n,k),n,k \right)\right]
									= \sum_k  \left[ s\left( \hat{t}_0 (\textbf{r}^\prime,n,k),n,k\right) \right] \\
								  &	= \sum_k  \lbrace G\left((\hat{t}_0-t_0)(\textbf{r}^\prime,n,k),\textbf{r}^\prime,k  \right) \cdot exp \left[ 2\pi i f_0 \left (\hat{t}_0-t_0\right)(\textbf{r}^\prime,n,k)  \right] \rbrace
\end{split}
\label{Equ:Theory_DASBeamforming}
\end{equation}
where $\hat{t}_0$ is the anticipated round-trip time for point $\textbf{r}$. Deviations of the true SoS $c(\textbf{r}^\prime)$ from the anticipated value $\hat{c}(\textbf{r}^\prime)$ lead to a deviation of the actual round-trip time from the anticipated value. This deviation, termed aberration delay, consists of two parts: one is the delay $\tau_{tx} (\textbf{r}^\prime,n)$ of the transmitted wave front when propagating towards $\textbf{r}^\prime$, and the other one is the delay  $\tau_{rx} (\textbf{r}^\prime,k)$ of echoes propagating from $\textbf{r}^\prime$ to element $k$, so that $t_0=\hat{t}_0 + \tau_{tx}+ \tau_{rx}$. With the aberration delays, Eq. \ref{Equ:Theory_DASBeamforming} ($\textbf{r}^\prime$ and $\textbf{r}$ are omitted for notational simplicity) becomes: 

\begin{equation}
u(n) = \sum_k \lbrace G(-\tau_{tx}(n) - \tau_{rx}(k),k) \cdot exp \left[ 2\pi i f_0 \left( -\tau_{tx}(n) - \tau_{rx}(k) \right) \right] \rbrace
\label{Equ:Theory_DASBeamformingAbberationDelay}
\end{equation}

\subsection*{The old model}
The old model was based on the following rationale: With the bandlimited frequency response of typical clinical US probes, $G$ varies ‘slowly’ compared to the oscillations of the exponential carrier. For a sufficiently small difference between Tx angles $\phi_n$ and $\phi_n ^\prime$, the change in $\tau_{tx}$ will be small compared to the temporal variation of $G$, so that the value of $G$ can be assumed constant and $\tau_{tx}$ can be replaced by its average $\overline{\tau_{tx}}$ in the envelope. Eq. \ref{Equ:Theory_DASBeamformingAbberationDelay} is then simplified to:

\begin{equation}
u(n) \simeq exp \left( -2\pi i f_0 \tau_{tx} (n)  \right) \sum_k \left[ G \left(-\overline{\tau_{tx}} - \tau_{rx} (k),k  \right) \cdot exp \left( -2\pi i f_0 \tau_{rx} (k)   \right) \right]
\label{Equ:Theory_RecSignalOldModel}
\end{equation}
On the right-hand side of Eq. \ref{Equ:Theory_RecSignalOldModel}, only the complex pre-factor depends on $n$. In the old model, the first processing step of CUTE is the experimental determination of the change in this pre-factor as a function of $n$. This is achieved by determining a map of local echo phase shift $\Theta(\textbf{r}^\prime, n, n^\prime)$, defined as the phase angle of the (locally averaged) point-wise Hermitian product between the crf-mode images obtained with Tx settings $n$ and $n^\prime$:


\begin{equation}
\Delta \Theta(\textbf{r}^\prime, n, n^\prime) \simeq arg \left[ \int_{-0.5\Delta x}^{0.5 \Delta x} \int_{-0.5 \Delta z}^{0.5\Delta z} d\boldsymbol{\hat{r}} \left\lbrace u(\textbf{r}^\prime + \boldsymbol{\hat{r}},n)\cdot u^\star (\textbf{r}^\prime + \boldsymbol{\hat{r}},n^\prime) \right\rbrace \right]
\label{Equ:Theory_PhaseShiftOldModel}
\end{equation}
Averaging over a limited area around $\textbf{r}^\prime$ (‘tracking kernel’), with size $\Delta x$ by $\Delta z$, improves robustness of the phase determination. The size of the tracking kernel defines the trade-off between spatial and contrast resolution of CUTE. Provided that the change in $\tau_{tx}$ is smaller than half the oscillation period (to avoid phase aliasing), the measured echo phase shift is - according to Eq. \ref{Equ:Theory_RecSignalOldModel} - related to the Tx aberration delay $\tau_{tx} (\textbf{r}^\prime, n)$, as:


\begin{equation}
\Delta \Theta(\textbf{r}^\prime, n, n^\prime) \simeq 2 \pi f_0 \Delta \tau_{tx}(\textbf{r}^\prime, n, n^\prime) = 2 \pi f_0	\left\lbrace \tau_{tx}(\textbf{r}^\prime,n^\prime) - \tau_{tx}(\textbf{r}^\prime,n) \right\rbrace
\label{Equ:Theory_PhaseShiftOldModelAberr}
\end{equation}

Similar to UCT, including US diffraction and refraction in the forward model is in principle feasible, but would result in a time-consuming iterative reconstruction. In view of real time SoS imaging, we thus adhere to the straight ray approximation and relate $\tau_{tx}(\textbf{r}^\prime,n)$ to line integrals of slowness deviation $\Delta \sigma(\textbf{r},n)$, from Tx elements $j_n$ to points $\textbf{r}^\prime$.


\begin{equation}
\tau_{tx}(\textbf{r}^\prime,n) = \int_{j_n}^{\textbf{r}^\prime} dl \left\lbrace \dfrac{1}{c(\textbf{r})} - \dfrac{1}{\hat{c}(\textbf{r})} \right\rbrace \equiv \int_{j_n}^{\textbf{r}^\prime} dl \Delta \sigma(\textbf{r})
\label{Equ:Theory_AberrationDelayStraightRayOldModel}
\end{equation} 

By inverting the forward model consisting of Eq. \ref{Equ:Theory_PhaseShiftOldModelAberr} and Eq. \ref{Equ:Theory_AberrationDelayStraightRayOldModel} the SoS $c$ is reconstructed from the measurements $\Delta \Theta$.

\subsection*{New model}
As mentioned in the Introduction, two fundamental changes to the CUTE methodology lead to the new model: In a first step, the common mid-angle (CMA) approach is described. Based on that, the new model is developed in a second step, which takes into account the position error of the reconstructed echoes.\\

\textbf{(a) Common mid-angle tracking}\\
An important prerequisite to the old model expressed in Eq. \ref{Equ:Theory_PhaseShiftOldModelAberr} was the assumption that $G$ did not depend on $n$. This assumption is plausible when considering a point reflector. In a more general situation, however, the relation between the aberration delay and the measured echo phase shift becomes ambiguous. This is illustrated for two extreme (but commonly found) cases: 

\begin{itemize}
\item Specular reflector: The echo from a reflector intersecting with $\textbf{r}^\prime$ propagates back to the probe along a main direction $\psi_n$ – determined by the mirror law – that varies when varying $\phi_n$. 
\item Uniform diffuse scattering: The echoes from a dense (below spatial resolution cell) and random distribution of ‘point’ reflectors interfere at the aperture. The signal detected by an element $k$ decorrelates with changing $n$, because the relative round-trip times of the different echoes change in a different way. The decorrelation rate depends on how many echoes interfere, i.e. on the \textit{a priori} unknown (due to aberrations) Tx focus diameter at $\textbf{r}^\prime$. 
\end{itemize}
In both cases – and also in a more general sense – the envelope function $G$ in Eq. \ref{Equ:Theory_ComplexSignalGenerated} depends on \textit{n}. In the same way as for the Tx angles, we assume sufficiently small differences between Rx angles so that the change in $\tau_{rx}$ will be small compared to the oscillation period. Then $\tau_{rx}$ can be replaced by its average $\overline{\tau_{rx}}$ in the envelope and Eq. \ref{Equ:Theory_RecSignalOldModel} can be re-written:

\begin{equation}
u(n) = exp\left(-2\pi i f_0 \tau_{tx} (n)\right) \sum_k \left\lbrace G\left( -\overline{\tau_{tx}} - \overline{\tau_{rx}},n,k \right)\cdot exp \left( -2\pi i f_0 \tau_{rx} (k) \right) \right\rbrace
\label{Equ:Theory_CommonMidAngle}
\end{equation}

\begin{figure}[h]
	\centering
	\includegraphics[width=1\textwidth]{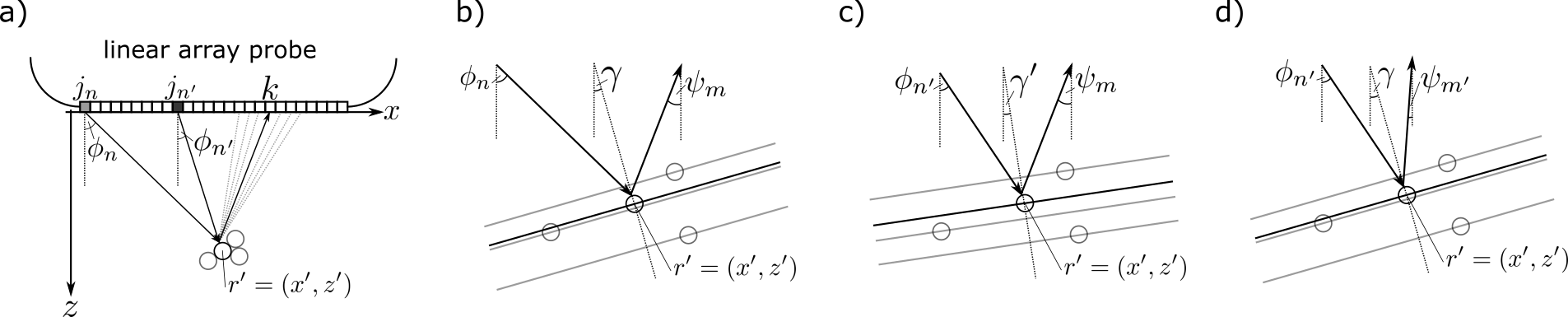}
\caption{a) Sketch of the echo US detection geometry. The US propagation paths are indicated leading from different elements $j_n$ towards a set of point reflectors at r' (circles) and back to an element k.  b-d) Zoom-up of the region around $\textbf{r}^\prime$, showing the isochrones from the point reflectors when detected under different pairs of Tx and Rx angles. Only when angles are grouped around the same mid-angle (b and d) the relative position of isochrones is unchanged, making robust tracking possible. }
\label{Fig:Theory1}
\end{figure}

Because $G$ varies with changing $n$, the sum in Eq. \ref{Equ:Theory_CommonMidAngle} depends on n. Thus, the phase shift according to Eq. \ref{Equ:Theory_PhaseShiftOldModel} not only depends on the Tx aberration delay but also on the Rx aberration delay in the complex exponential in the right-hand side of Eq. \ref{Equ:Theory_CommonMidAngle}. 
To avoid ambiguities, we introduce a fundamental change to the CUTE methodology: common mid-angle (CMA) tracking. In this technique, crf-mode images are reconstructed for pairs of Tx angles $\phi_n$ and Rx angles $\psi_m$ that have the same mid-angle. One has to keep in mind that Rx beamsteering goes hand-in-hand with reducing the Rx angular aperture, at the cost of a reduced lateral resolution of the crf-mode image. CUTE is based on evaluating echo phase shifts in the beamformed crf-mode image as opposed to the channel data, specifically because this allows to spatially resolve the influence of the SoS on the echo phase shift of different echoes. When reconstructing images with sharp $\phi_n$ and $\psi_m$, no lateral resolution would be obtained in the crf-mode images. Therefore, to achieve lateral resolution, images have to be reconstructed with non-zero Tx and Rx angular apertures, which in combination define the size of a spatial resolution cell. Even though, experimentally, angles have to be defined with non-zero angular aperture, we continue assuming that they are sharply defined for the remainder of this chapter. Later on, we will discuss the practical implication of the non-zero aperture on SoS reconstruction.  
Similar to the common mid-point approach that was previously used in aberration correction \cite{jaeger2015full}, the CMA approach makes use of signal redundancy: Signals obtained under angle pairs $(\phi_n, \psi_m)$ grouped around the same mid-angle $\gamma =  0.5 \left( \phi_n+\psi_m \right)$ are well correlated, independent of the spatial distribution of US reflectors. For illustration, Fig. \ref{Fig:Theory1}(b) shows a scale-up of a spatial resolution cell centred around point $\textbf{r}^\prime$, containing a number of reflectors. Each reflector is the pivot of an isochrone, a line that connects all points that would lead – for a specific combination of $(\phi_n, \psi_m)$ - to the detection of an echo at the same time. When changing $\phi_n$ to $\phi_{n^\prime}$ without changing $\psi_m$ (Fig. \ref{Fig:Theory1}(c)), the distances between the isochrones change, leading to decorrelation of the signal that is detected by elements located in direction $\psi_m$. When changing $\psi_m$ in opposite direction (Fig \ref{Fig:Theory1}(d)), to $\psi_m^\prime$ such that the mid-angle in the pairs $(\phi_n, \psi_m)$ and $(\phi_{n^\prime}, \psi_{m^\prime})$ is fixed (so that $\gamma = 0.5(\phi_n + \psi_m) = 0.5(\phi_{n^\prime} + \psi_{m^\prime}))$, then the distances between isochrones is not altered. 
 
We thus assume that crf-mode images $u(\textbf{r},n,m)$  are reconstructed where $n$ indexes the Tx setting resulting in a Tx angle $\phi_n$, $m$ indexes the mid-angle setting $\gamma_m$, and both together result in a receive angle $\psi_{(m,n)}$. 
Accordingly, Eq. \ref{Equ:Theory_PhaseShiftOldModel} and Eq. \ref{Equ:Theory_PhaseShiftOldModelAberr} are adapted for the CMA tracking to: 

\begin{equation}
\Delta \Theta(\textbf{r}^\prime,n,n^\prime,m) = arg \left[ \int_{-0.5\Delta x}^{0.5 \Delta x} \int_{-0.5 \Delta z}^{0.5 \Delta z} d\boldsymbol{\hat{r}} \left( u(\textbf{r}^\prime + \boldsymbol{\hat{r}},n,m)\cdot u^\star (\textbf{r}^\prime +\boldsymbol{\hat{r}},n^\prime,m) \right) \right]
\label{Equ:Theory_PhaseShiftCMAIntegrals}
\end{equation}
\begin{equation}
\begin{split}
\Delta \Theta(\textbf{r}^\prime,n,n^\prime,m) 	& = 2 \pi f_0 \left( \Delta \tau_{tx}(\textbf{r}^\prime, n , n^\prime) + \Delta  \tau_{rx}(\textbf{r}^\prime, m, n , n^\prime) \right) \\
												& = 2 \pi f_0 \left( \tau_{tx} (\textbf{r}^\prime,n^\prime) - \tau_{tx}(\textbf{r}^\prime,n) + \tau_{tr}(\textbf{r}^\prime,m,n) - \tau_{tr}(\textbf{r}^\prime, m , n^\prime)   \right)
\end{split}
\label{Equ:Theory_PhaseShiftCMA}
\end{equation}

\begin{figure}[h]
	\centering
	\includegraphics[width=1\textwidth]{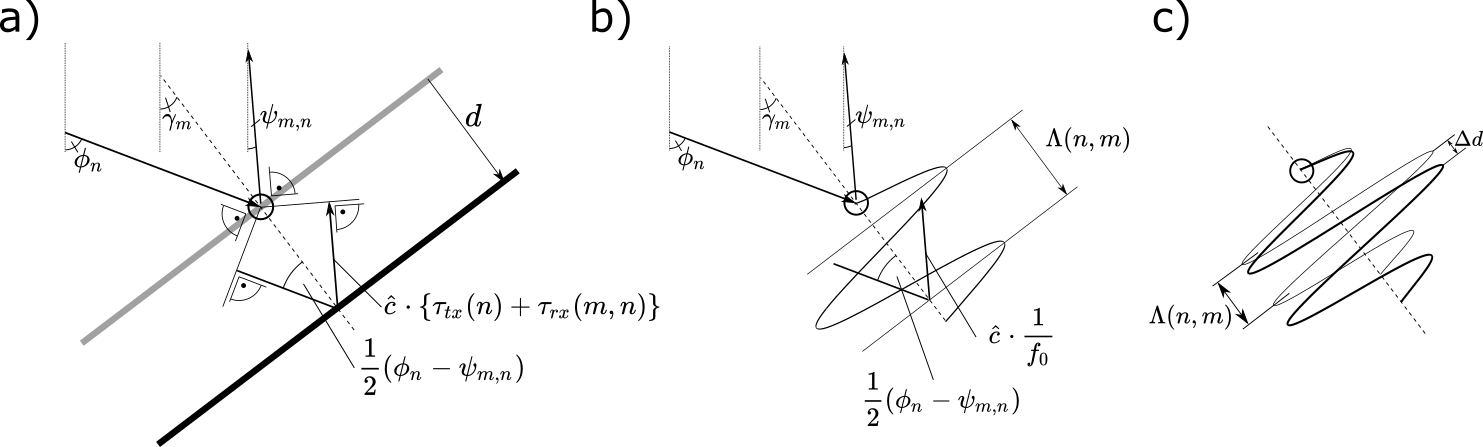}
\caption{Geometrical sketch showing the relation between the different angles, the aberration delays and the spatial shift of an isochrone. b) Similar sketch showing the relation between the centre frequency and the echo’s spatial oscillation period. c) The measured phase shift is determined by the shift $\Delta d$ of the centroid of the echo envelope.}
\label{Fig:Theory_NewModel}
\end{figure}

\textbf{(b) Reconstructed echo position error model}\\
To derive the second fundamental change to the CUTE methodology, we introduce a geometric perspective of how aberration delays relate to the echo phase (Fig. \ref{Fig:Theory_NewModel}). 

We assume a reflector (point or plane) that is detected with a specific combination $(\phi_n,\psi_{m,n})$. The total aberration delay (in Fig. \ref{Fig:Theory_NewModel}(a) larger than zero) leads to a shift of the reconstructed position of the echo, away from the true location of the reflector, to the isochrone for which the anticipated round-trip time agrees with the actual round-trip time. The distance $d$ between the location of the reflector and the isochrone is – according to the geometric considerations depicted in Fig. \ref{Fig:Theory_NewModel}(a) – given by: 

\begin{equation}
d(n,m) =  \dfrac{\hat{c} \left[ \tau_{tx}(n) + \tau_{rx}(m,n) \right]} {2\cos\left[\frac{1}{2}\left( \phi_n - \psi_{m,n} \right) \right] }
\label{Equ:TheoryNew_d}
\end{equation}

To determine the phase shift, the spatial oscillation period of the reconstructed echo is also required. This follows the same argumentation (see Fig. \ref{Fig:Theory_NewModel}(b)): The spatial period $\Lambda$ is given by the distance between two isochrones that are separated by one period in time: 

\begin{equation}
\Lambda(n,m) = \dfrac{\frac{\hat{c}}{f_0}}{2\cos\left[\frac{1}{2}\left( \phi_n - \psi_{m,n} \right) \right] }
\label{Equ:TheoryNew_Lamda}
\end{equation}

An important prerequisite of the old model was the evaluation of the echo phase shift at the true location of the ultrasound reflector ($\textbf{r}$ was chosen equal to $\textbf{r}^\prime$). 


However, it is not the phase shift at the true location of the reflector that determines the value of the measured phase shift, but the shift of the reconstructed position of the echo. As a second fundamental modification to the CUTE methodology, this is taken into account by the new model. Fig. \ref{Fig:Theory_NewModel}(c) illustrates that the phase shift is given by the ratio between  $\Delta d$, the shift of the echo centroid and $\Lambda$. Formally, when changing $\phi_n$ by $\Delta \phi$ to $\phi_{n^\prime}$ (so that $\Psi_{m,n}$ changes by $-\Delta \phi$ according to $\Psi_{m,n^\prime}$), $d$ changes accordingly by: 

\begin{equation}
\Delta d = \dfrac{\hat{c}\left[ \tau_{tx}(\phi_n + \Delta \phi) + \tau_{rx}(\psi_{m,n} - \Delta \phi) \right]}  {2 \cos \left[\frac{1}{2}\left(\phi_n - \psi_{m,n} + 2 \Delta \phi \right) \right]} - \dfrac{\hat{c} \left[ \tau_{tx}(\phi_n) + \tau_{rx}(\psi_{m,n}) \right] }{2 \cos\left[\frac{1}{2} \left(  \phi_n - \psi_{m,n}\right)\right]}
\end{equation}

To simplify the maths, we assume that the signals are bandpass filtered in a way that the spatial period $\Lambda$ does not change between combinations $(n, m)$ and $(n^\prime, m^\prime)$ (this can easily be achieved in practise) and takes, for example, the value: 

\begin{equation}
\Lambda = \dfrac{\frac{\hat{c}}{f_0}}{2}
\label{Equ:Theory_NewModelLambda}
\end{equation}

With this convention, the new model for the phase shift then is:

\begin{equation}
\begin{split}
\Delta \Theta (\textbf{r}^\prime,n,n^\prime,m) &= \dfrac{2 \pi \Delta d}{\Lambda} \\ &= 2\pi f_0 \left\lbrace \dfrac{\tau_{tx}(n) + \tau_{rx}(m,n^\prime)}{\cos\left[ \frac{1}{2}\left( \phi_{n^\prime} - \psi_{m,n^\prime} \right) \right]}  - \dfrac{\tau_{tx}(n) + \tau_{rx}(m,n)}{\cos\left[ \phi_n - \psi_m \right]} \right\rbrace\\
 &= 2 \pi f_0 \left\lbrace \dfrac{\tau_{tx}(\phi_n + \Delta \phi) + \tau_{rx} (\psi_{m,n} - \Delta \phi)}{ \cos \left[ \frac{1}{2} \left( \phi_n - \psi_{m,n} + 2\Delta \phi \right) \right]} - \dfrac{\tau_{tx}(\phi_n) + \tau_{tx}(\psi_{m,n})}{\cos \left[ \frac{1}{2} \left(\phi_n - \psi_{m,n} \right) \right]} \right\rbrace
\end{split}
\label{Equ:Theory_NewModelPhaseShift}
\end{equation}

In comparison to the old model, the new model (eq. \ref{Equ:Theory_NewModelPhaseShift}) includes CMA tracking and a division of the aberration delay by the different cosines for the different ($n$, $n^\prime$).

\section{Materials and Methods}
\subsubsection*{Space domain forward and inverse model}
For simplicity of formulation of the numerical models and without loss of generality, we have chosen a plane wave pulse-echo acquisition approach so that each pixel in the image area is insonified with the same discrete set $\lbrace\phi_n\rbrace$ of equidistantly spaced Tx angles. The equidistant spacing of the Tx angles allows choosing – for the CMA approach – a set $\lbrace\gamma_m \rbrace_n$ of common mid-angles that result in a set $\lbrace\psi_m \rbrace_n$ of Rx angles that is identical to $\lbrace\phi_n\rbrace$. This allows to make use of data redundancies as explained below. For clarity, we introduce a new notation for the angle set, $\lbrace \Phi_l\rbrace$, with $\phi_n \in \lbrace \Phi_l\rbrace $  and $\psi_{m,n} \in \lbrace \Phi_l \rbrace$. For this study, $\lbrace \Phi_l \rbrace$ was chosen to be [-25$^\circ$, -15$^\circ$, -5$^\circ$, 5$^\circ$, 15$^\circ$, 25$^\circ$] for $l$ = [1 .. 6]. Table \ref{Tbl:Materials_Table1} summarizes the resulting combinations of $\phi$, $\psi$ and $\gamma$. As described in the Theory section, tracking can only be performed between angle combinations having the same mid-angle $\gamma$, indicated in table \ref{Tbl:Materials_Table1} by fields having the same gray value. Since only one combination of $\phi$, $\psi$ exists for both common mid angles $\gamma = \pm 25^\circ$, no tracking was performed for these $\gamma$.

\begin{table}[ht]
\centering
\begin{tabular}{|c|ccccccc|}
\hline
$\psi_{m,n}$ & $\phi_n$                                          & -25                                                & -15                                              & -5                          & 5                           & 15                         & 25                                                \\ \hline
-25 & \multicolumn{1}{c|}{}                        & \cellcolor[HTML]{343434}{\color[HTML]{FFFFFF} -25} & \cellcolor[HTML]{656565}-20                      & \cellcolor[HTML]{9B9B9B}-15 & \cellcolor[HTML]{C0C0C0}-10 & \cellcolor[HTML]{D9D9D9}-5 & 0                                                 \\
-15 & \multicolumn{1}{c|}{}                        & \cellcolor[HTML]{656565}{\color[HTML]{000000} -20} & \cellcolor[HTML]{9B9B9B}-15                      & \cellcolor[HTML]{C0C0C0}-10 & \cellcolor[HTML]{D9D9D9}-5  & 0                          & \cellcolor[HTML]{D9D9D9}5                         \\
-5  & \multicolumn{1}{c|}{}                        & \cellcolor[HTML]{9B9B9B}-15                        & \cellcolor[HTML]{C0C0C0}-10                      & \cellcolor[HTML]{D9D9D9}-5  & 0                           & \cellcolor[HTML]{D9D9D9}5  & \cellcolor[HTML]{C0C0C0}10                        \\
5   & \multicolumn{1}{c|}{}                        & \cellcolor[HTML]{C0C0C0}-10                        & \cellcolor[HTML]{D9D9D9}-5                       & 0                           & \cellcolor[HTML]{D9D9D9}5   & \cellcolor[HTML]{C0C0C0}10 & \cellcolor[HTML]{9B9B9B}15                        \\
15  & \multicolumn{1}{c|}{}                        & \cellcolor[HTML]{D9D9D9}-5                         & 0                                                & \cellcolor[HTML]{D9D9D9}5   & \cellcolor[HTML]{C0C0C0}10  & \cellcolor[HTML]{9B9B9B}15 & \cellcolor[HTML]{656565}20                        \\
25  & \multicolumn{1}{c|}{\multirow{-6}{*}{$\gamma$}} & 0                                                  & \cellcolor[HTML]{D9D9D9}{\color[HTML]{000000} 5} & \cellcolor[HTML]{C0C0C0}10  & \cellcolor[HTML]{9B9B9B}15  & \cellcolor[HTML]{656565}20 & \cellcolor[HTML]{343434}{\color[HTML]{FFFFFF} 25} \\ \hline
\end{tabular}
 \caption{Combinations of Tx ($\phi_n)$, Rx ($\psi_{m,n}$) and common mid-angles ($\gamma$) used in this study (in $^\circ$). Tracking is performed between angle combinations having the same mid-angle $\gamma$, indicated by fields having the same gray value. No tracking was performed for $\gamma$ = $\pm$25$^\circ$.}
\label{Tbl:Materials_Table1}
\end{table}

Choosing identical sets of Tx and Rx angles allows making use of data redundancy: In an ideal case, the crf-mode images resulting from interchangeable $\left( \phi_n \mid \psi_{m,n} \right)$ angle pairs  $\left( \Phi_l \mid \Phi_{l^\prime} \right)$ and $\left( \Phi_{l^\prime} \mid \Phi_l \right)$ are identical by time reversal symmetry. 
The phase correlation between $\left( \Phi_l \mid \Phi_{l^\prime} \right)$ and $\left( \Phi_{l+1} \mid \Phi_{l^\prime-1} \right)$ and the one between $\left( \Phi_l \mid \Phi_{l^\prime} \right)$ and $\left( \Phi_{l^\prime-1} \mid \Phi_{l+1} \right)$ thus provide redundant data. In a more realistic case, slight differences exists between Tx and Rx beamforming. This asymmetry is removed by averaging of the redundant phase correlations. In addition, the averaging results in phase shift values that are more robust against phase aliasing, and the reduced total number of phase shift maps allows a faster SoS reconstruction. \\

Depending on the Tx angle, the Tx wave front reaches only a part of the imaging plane, leaving areas of missing echoes ('Tx shadows') towards the edges of the image. As mentioned in the Introduction, the previously proposed FD formulation cannot account for missing data regions, which complicates a quantitative SoS reconstruction. To avoid such drawbacks, the forward models in this study were implemented in space domain (SD). For this purpose, the measured spatial distribution of echo phase shift as well as the distribution of slowness to be reconstructed were discretised on the same 2D Cartesian grid, as 

\begin{align}
\Delta \sigma(\textbf{r}^\prime) 				& \rightarrow \Delta \sigma_{j,k}  \nonumber \\
\Delta \Theta(\textbf{r}^\prime,n,n^\prime) 	& \rightarrow \Delta \Theta_{j,k}(n,n^\prime) 	&  \textnormal{(classical tracking)}\\
\Delta \Theta(\textbf{r}^\prime,n,n^\prime,m) 	& \rightarrow \Delta \Theta_{j,k}(n,n^\prime,m) & \textnormal{(CMA tracking)} \nonumber
\end{align}

Thereby, the number of nodes $(j,k)$ in $x-$ and $z-$ direction is $N_j$ and $N_k$, respectively. The nodes sample the chosen dimensions $(X,Z)$ of a rectangular image area (in our study 38.4 by 50 mm) with a spatial resolution of $(\Delta x,\Delta z)$ (in our study 0.96 by 1 mm).\\

The goal of the experimental part is to compare the old and the new model in a phantom study. Thus, both the old and the new model were implemented as described by Eq. \ref{Equ:Materials_OldModel} and  \ref{Equ:Materials_NewModel}, respectively, derived from Eq. \ref{Equ:Theory_PhaseShiftOldModelAberr}, \ref{Equ:Theory_AberrationDelayStraightRayOldModel} and \ref{Equ:Theory_NewModelPhaseShift}. However, to demonstrate that the improvement of the new model not only depends on the CMA tracking, we also implemented a forward model where only CMA tracking - but not the echo position error - is taken into account (Eq. \ref{Equ:Materials_CMAModel}).



\begin{itemize}
\item Old model
\begin{align}
& \Delta \Theta_{j,k}(n,n^\prime) = \nonumber \\
& 2 \pi f_0 \left[ \sum_{j^\prime,k^\prime} w_{(j,k,j^\prime, k^\prime)}^{\phi_{n^\prime}} \Delta \sigma_{(j^\prime,k^\prime)} - \sum_{j^\prime,k^\prime} w_{(j,k,j^\prime, k^\prime)}^{\phi_n} \Delta \sigma_{(j^\prime,k^\prime)} \right]
\label{Equ:Materials_OldModel}
\end{align}

\item CMA model

\begin{align}
& \Delta \Theta_{j,k}(n,n^\prime,m) = \nonumber \\
& 2 \pi f_0 \left[ \left( \sum_{j^\prime,k^\prime} w_{(j,k,j^\prime, k^\prime)}^{\phi_{n^\prime}} \Delta \sigma_{(j^\prime,k^\prime)} + \sum_{j^\prime,k^\prime} w_{(j,k,j^\prime, k^\prime)}^{\psi_{n^\prime,m}} \Delta \sigma_{(j^\prime,k^\prime)}  \right) \right. -  \label{Equ:Materials_CMAModel} \\
& \left. \left( \sum_{j^\prime,k^\prime} w_{(j,k,j^\prime, k^\prime)}^{\phi_n} \sigma_{(j^\prime,k^\prime)} + \sum_{j^\prime,k^\prime} w_{(j,k,j^\prime, k^\prime)}^{\psi_{n,m}} \sigma_{(j^\prime,k^\prime)} \right) \right] \nonumber
\end{align}
\item New model
\begin{align}
& \Delta \Theta_{j,k}(n,n^\prime,m) = \nonumber \\ 
& \dfrac{2\pi f_0}{\cos \left[\frac{1}{2}\left(\phi_{n^\prime} - \psi_{n^\prime,m} \right) \right]} \left( \sum_{j^\prime,k^\prime} w_{(j,k,j^\prime, k^\prime)}^{\phi_{n^\prime}} \Delta \sigma_{(j^\prime,k^\prime)} + \sum_{j^\prime,k^\prime} w_{(j,k,j^\prime, k^\prime)}^{\psi_{n^\prime,m}} \Delta \sigma_{(j^\prime,k^\prime)}  \right) - \label{Equ:Materials_NewModel}  \\ 
& \dfrac{2\pi f_0}{\cos \left[\frac{1}{2}\left(\phi_{n} - \psi_{n,m} \right) \right]} \left( \sum_{j^\prime,k^\prime} w_{(j,k,j^ \prime, k^\prime)}^{\phi_n} \Delta \sigma_{(j^\prime,k^\prime)} + \sum_{j^\prime,k^\prime} w_{(j,k,j^\prime, k^\prime)}^{\psi_{n,m}} \Delta\sigma_{(j^\prime,k^\prime)} \right) \nonumber
\end{align}

\end{itemize}
Thereby the $w_{(j,k,j^\prime, k^\prime)}^{\Phi_l} $ are the integration weights defining the discrete line integrals along the US propagation direction with the angles $\Phi_l$.\\
Any of the forward models in Eq. \ref{Equ:Materials_OldModel} to \ref{Equ:Materials_NewModel}  can be written in matrix notation: 
\begin{equation}
\Delta \boldsymbol{\theta} = \boldsymbol{M} \Delta \boldsymbol{\sigma}
\label{Equ:Materials_ForwardMatrixFormulation}
\end{equation}
To reconstruct the values $\Delta \sigma_{j^\prime, k^\prime}$ from the measurements $\Delta \theta_{j,k}$, a Tikhonov pseudo-inverse of $\boldsymbol{M}$ is used: 

\begin{align}
\Delta \boldsymbol{\sigma} &= \boldsymbol{M}^{\textnormal{inv}} \Delta \boldsymbol{\theta} \label{Equ:Materials_InverseMatrixFormulation}\\
\boldsymbol{M}^{\textnormal{inv}} &= \left(\boldsymbol{M}^T \boldsymbol{M} + \gamma_x \boldsymbol{D}_x^T\boldsymbol{D}_x + \gamma_z \boldsymbol{D}_z^T\boldsymbol{D}_z\right)^{\textnormal{inv}}\boldsymbol{M}^T \label{Equ:Materials_InverseMatrixFullFormulation}
\end{align}
which minimizes the expression
\begin{align*}
C(\Delta \sigma) = \Vert \Delta \boldsymbol{\theta} - \boldsymbol{M} \Delta \boldsymbol{\sigma} \Vert^2_2 + \Vert  \gamma_x \boldsymbol{D}_x \Delta \boldsymbol{\sigma} +  \gamma_z \boldsymbol{D}_z \Delta \boldsymbol{\sigma} \Vert^2_2
\end{align*}

Thereby, $\boldsymbol{D}_x$ and $\boldsymbol{D}_z$ are finite difference operators in $x$ and $z$, respectively, and $\gamma_x$ and $\gamma_z$ are regularisation parameters. Regularisation of the spatial gradient of the slowness deviation enforces a smooth slowness profile while letting the mean SoS vary freely. 

\subsubsection*{Data acquisition system}
For the experimental study, we used a Vantage 64 LE (Verasonics Inc., WA, USA). This research ultrasound system allows simultaneous ultrasound transmission on 128 channels and parallel digitisation of signals on 64 elements, with real time data transfer via a PCI Express link to a host computer for further processing. The system was connected with an L7-4 linear vascular probe (ATL Philips, WA, USA) for pulse-echo signal acquisition. This probe features 128 elements at a 0.29 mm pitch (resulting in 38.4 mm aperture length), and a bandwidth from 4 to 7 MHz with 5 MHz centre frequency. 
We implemented a dedicated scan sequence for the acquisition of plane wave pulse-echo data. As mentioned earlier, the target Tx and Rx steering angle step size was 10$^\circ$. Given the SoS contrast of the chosen phantoms, however, this step size resulted in phase aliasing due to aberration delay changes above half the oscillation period at the centre frequency. To avoid phase aliasing, we chose a tracking angle step size of 2$^\circ$. Moreover, for synthetic Tx focusing (via coherent plane wave compounding \cite{bercoff2011ultrafast, montaldo2009coherent}), data had to be acquired with an even smaller step size of 0.5$^\circ$. Plane wave data was thus acquired with Tx angles $\phi$ ranging from -27.5$^\circ$ to 27.5$^\circ$ in 0.5$^\circ$ steps. To use the full probe aperture on receive, echoes were recorded twice for each Tx angle, once on each of two 64-element sub-apertures.

\subsubsection*{CRF-mode image reconstruction}
\label{SubSubSec:CRF-Mode}
Crf-mode images were reconstructed off-line for each Tx angle from the raw crf data using a delay-and-sum (DAS) algorithm. Crf-mode amplitudes were calculated on a rectangular grid, with dimensions 38.4 mm in $x$ (aperture length) by 50 mm in $z$ direction and with spatial resolution $dx = 0.29$ mm in $x$ (pitch) by $dz = 0.037$ mm in z. At each grid node with coordinates $(x,z)$, the anticipated round-trip times $\hat{t}_0(x,z,\phi,k)$ to each receiving element k were calculated based on the anticipated SoS $\hat{c}$. 

\begin{align}
u(x,z,\phi) &= \sum_k\left[\alpha(x,z,k) \cdot s\left( \hat{t}_0(x,z,\phi,k)\right)\right]\\
\hat{t}_0(x,z,\phi,k) &= \dfrac{1}{\hat{c}} \left[ \sin(\phi) \left(x-\tan(\phi)z \right)+\cos(\phi)z + \sqrt{\left(k-x\right)^2 + z^2} \right]
\label{Equ:Materials_DASEquation}
\end{align}
The weights $\alpha(x,z,k)$ were designed to provide (within the limits of the aperture length) a receive angular aperture of $\pm$ 30$^\circ$. \\

For focused Tx steering along each specific target angle $\phi$, crf-mode images were coherently compounded over a Tx angle range $[\phi-2.5^\circ:0.5^\circ:\phi+2.5^\circ]$ centred around the respective target angle, resulting in a synthetic Tx focus with an angular aperture of 5$^\circ$. This was done for a target Tx angle range $\phi \in [-25^\circ:2^\circ:25^\circ]$.\\

For the CMA approach, each crf-mode image was transformed to a sequence of crf-mode images for different receive steering angles $\Psi \in \left[-25^\circ:2^\circ:25^\circ \right]$. This was achieved by spatial filtering of the already reconstructed crf-mode images. For this purpose, the 2D discrete Fourier transform was first calculated. Then the spatial frequency spectrum was multiplied with an angle-dependent weighting function to highlight a specific mid-angle, in that way implicitly determining the corresponding receive angle. The angular radius of the weighting function was chosen $\pm$ 1.25$^\circ$, resulting in an implicit Rx angular aperture radius of 2.5$^\circ$. This approach was chosen, as opposed to an explicit Rx beamsteering, to avoid abundant memory usage and it was implemented as part of the tracking algorithm. \\

\subsubsection*{Phase tracking}

The echo phase shift was determined according to Eq. \ref{Equ:Theory_PhaseShiftOldModelAberr} (old model) and Eq. \ref{Equ:Theory_PhaseShiftCMAIntegrals} (CMA approach). The tracking kernel size was chosen $\Delta x$ = 2 mm and $\Delta z$ = 2 mm. Echo phase tracking was performed with the 2$^\circ$ angle steps and phase shift maps were accumulated over successive angle steps to obtain the echo phase shift over 10$^\circ$ steps between Tx angles $\Phi_l$ and $\Phi_{l+1}$ (conventional tracking) and Tx/Rx angle pairs $(\Phi_l \vert \Phi_{l^\prime})$ and $(\Phi_{l+1} \vert \Phi_{l^\prime-1})$ (CMA approach). 

\subsubsection*{SoS reconstruction implementation details}

For SoS reconstruction, the phase shift maps were downsampled onto a Cartesian grid with 40 $(x)$ by 50 $(z)$ pixels covering 38.4 mm $(x)$ by 50 mm $(z)$. The forward models according to Eq. \ref{Equ:Materials_ForwardMatrixFormulation} and \ref{Equ:Materials_InverseMatrixFormulation} were correspondingly formulated for a SoS distribution sampled on the same grid, resulting in square system matrices with (40$\cdot$50) by (40$\cdot$50) elements per angle step. The reconstructed SoS images, however, showed artefacts in the top 5 mm due to element cross talk. Further, artefacts occurred towards the lower edge of the image, independent of the initial choice of the depth range (reason yet unclear). To mask out these artefacts, the SoS images were cut in z-direction to a depth range from 5 to 40 mm for the final presentation. 

The regularization parameters $\gamma_x$ and $\gamma_z$ (see. Equ. \ref{Equ:Materials_InverseMatrixFullFormulation}) are subject to a trade-off between reducing artefacts (by enforcing a smooth slowness profile) and maximising spatial resolution. For this study, they were chosen so as to  clearly distinguish the different phantom compartments.


\subsubsection*{Phantom design and materials}
\begin{figure}[!htbp]
    \centering
	\includegraphics[width=1\textwidth]{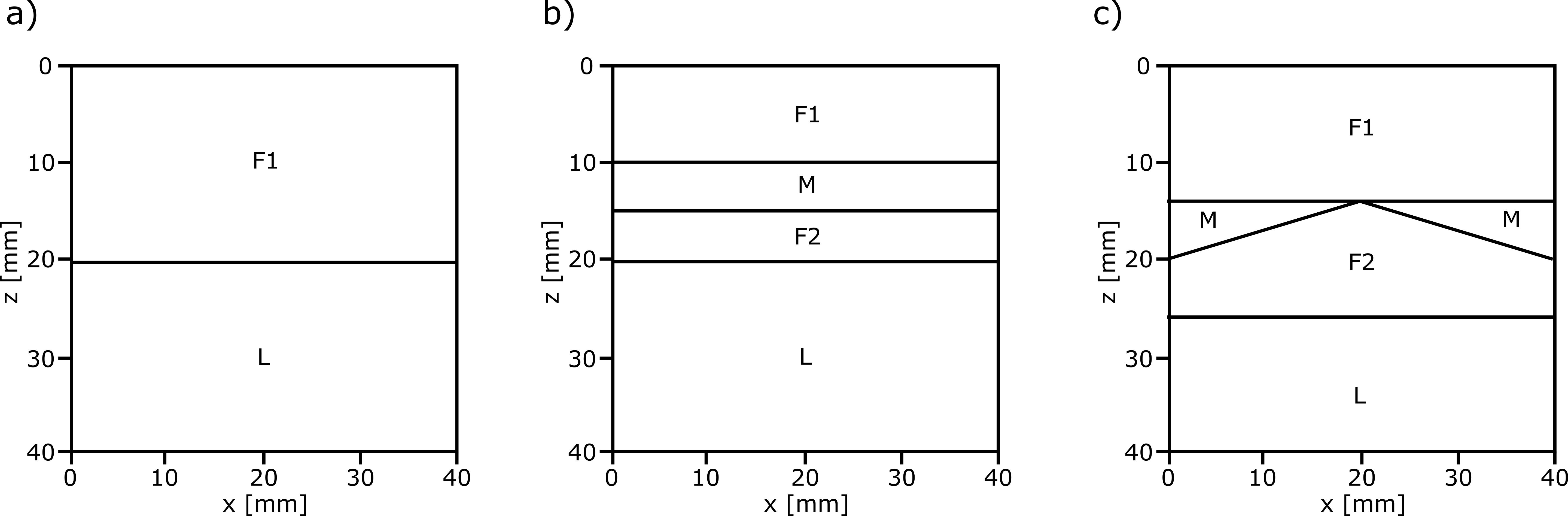}
	\caption{Sketches of the phantom geometries: (a) Two layer phantom mimicking the liver (L) and a single fat layer (F1); (b) Four layer phantom mimicking the liver (L), the post peritoneal fat layer (F2), the rectus abdominis (M) and the  subcutaneous fat layer (F1); (c) LVMD phantom mimicking the liver (L) covered by a triangular shaped fat layer (F2), rectus abdominis (M) and the subcutaneous fat layer (F1)}
	\label{Fig:Results_PhantomSketch}
\end{figure}

For this study, three different phantoms where designed, each representing as close as possible a particular geometry of the anatomical structure of the abdominal wall and the liver (see Fig. \ref{Fig:Results_PhantomSketch}).

\begin{enumerate}[label=(\alph*)]
\item \textit{Two layer phantom}: This phantom contained two horizontal (parallel to $x$) layers, mimicking a single fat layer (F1) on top of liver tissue (L). The special feature of this type is the complete absence of lateral SoS variations but a pronounced axial variation. This geometry occurs when imaging the liver sagittally through the linea alba.

\item \textit{Four layer phantom}: In comparison to the two layer phantom, two additional layers were added, mimicking the rectus abdominis muscle (M) and the post peritoneal fat layer (F2). The purpose of this phantom was to evaluate the ability of the different models in resolving axial SoS variations of thin layers, representative for a sagittal section lateral to the linea alba.   

\item \textit{Laterally varying muscle diameter (LVMD) phantom}: In contrast to the four layer phantom, the rectus abdominis (M) layer deviates in this phantom from the parallel layer structure and consists of two wedge-shaped areas, as when imaging the liver in a transverse section through the linea alba.

\end{enumerate}

The phantom components were based on oil-in-gelatin emulsions \cite{madsen1982anthropomorphic, madsen2003tissue,  madsen2006anthropomorphic, nguyen2014development}. In a first step, a gelatin base solution was prepared by dissolving 20 wt\% porcine gelatin (Geistlich Spezial Gelatine, health and life AG, Switzerland) in 75$^\circ$C H$_2$O and adding 2 wt\% cellulose (Sigmacell Cellulose Type 20, Sigma Aldrich, Switzerland) to provide uniform diffuse echogenicity. This aqueous gelatin base solution had a SoS of 1555 m/s and was used to mimic muscle (M) and liver (L). To mimic fat, medium-chain triglycerides oil (Ceres-MCT Oil, Puravita, Switzerland) (SoS = 1350 m/s) was slowly blended under continuous stirring into the aqueous gelatine base solution using a Visco Jet cone-stirrer (VISCO JET Agitation Systems, Germany). During this process, small oil droplets were formed and captured via hydrophobic interaction by the lipophilic part of the gelatine strings. After cooling the emulsion, the oil droplets were trapped within the gelatine matrix. The resulting SoS was determined by the emulsion's relative MCT oil weight content, in this study 0.65, resulting in a SoS of 1420 m/s. The mentioned SoS values were determined using a through-transmission time-of-fight set-up (accuracy $\pm$ 5 m/s) and will serve in the following as reference values for the SoS images. Note that, even though the absolute SoS values may deviate from real tissue depending on literature references, we have taken care to chose a representative SoS contrast.

\newpage
\section{Phantom Results}
\begin{figure}[!htbp]
    \centering
    \textbf{Two layer phantom}\par\medskip
	\includegraphics[width=1\textwidth]{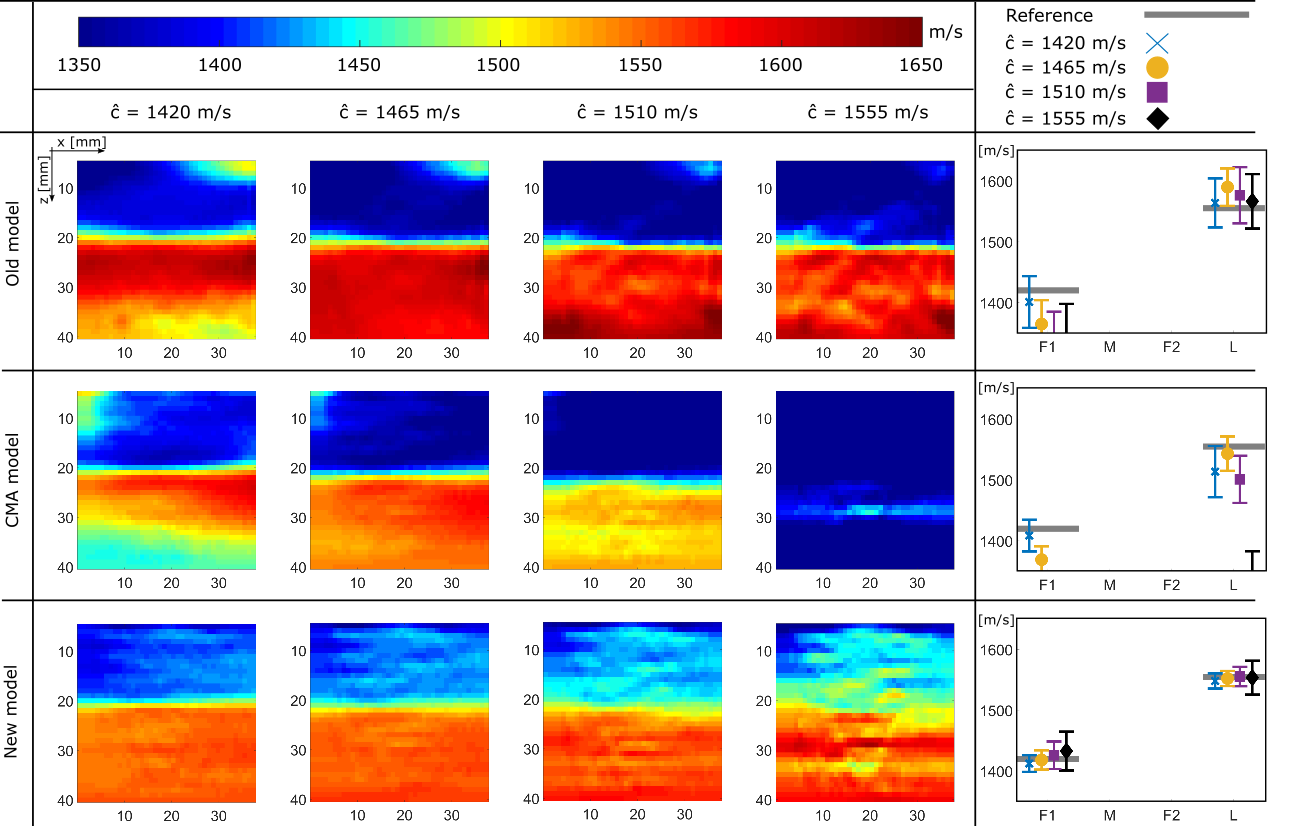}
	\caption{SoS images of the two layer phantom mimicking the liver (SoS = 1555 m/s) and a single fat layer (SoS = 1420 m/s) (from posterior to anterior). The SoS images were reconstructed based on the different forward models (rows) and four different \textit{a priori} SoS values $\hat{c}$ (columns). }
	\label{Fig:Results_2LayerSoSImage140ms}
\end{figure}

\begin{figure}[!htbp]
    \centering
    \textbf{Four layer phantom}\par\medskip
	\includegraphics[width=1\textwidth]{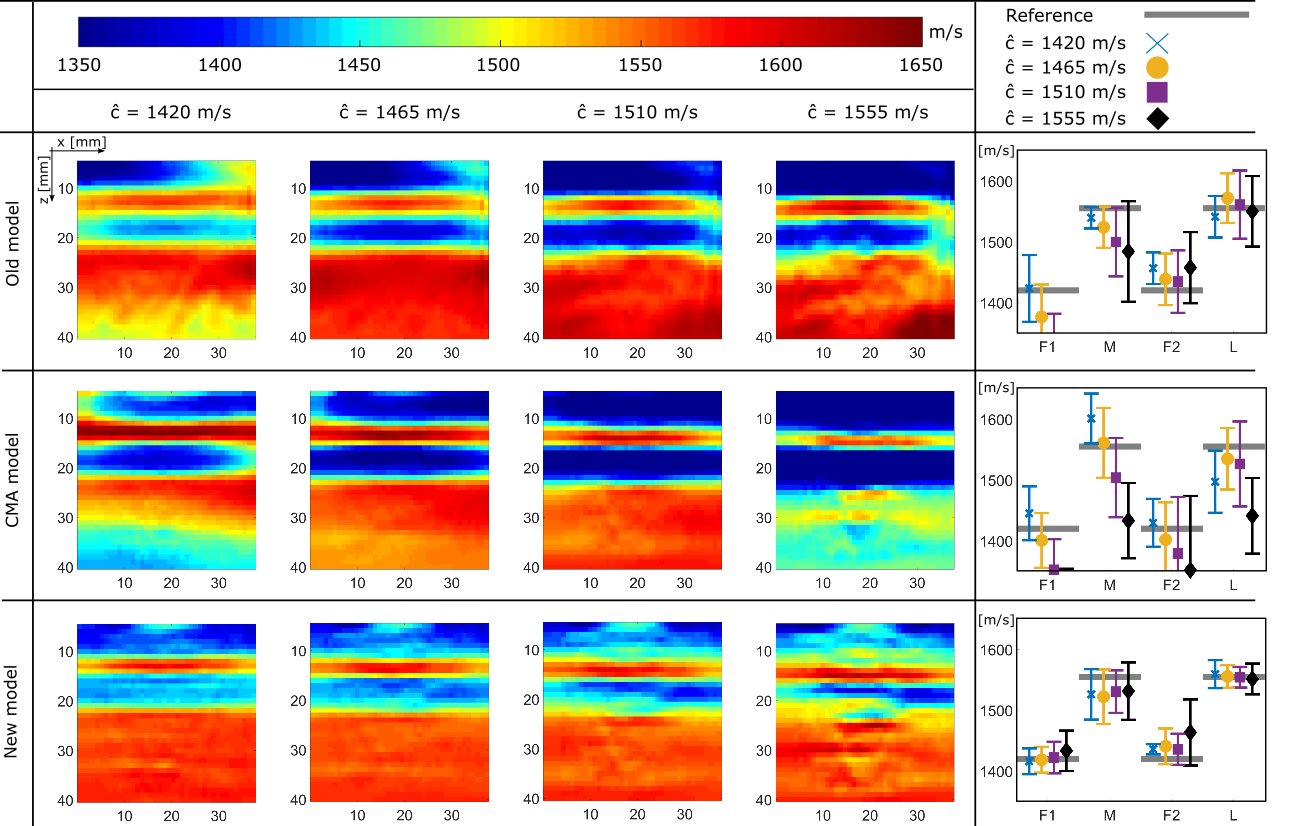}
	\caption{SoS images of the four layer phantom mimicking the liver (SoS = 1555 m/s), a thin fat layer (SoS = 1420 m/s), the rectus abdomins (SoS = 1555 m/s) and another thin fat layer (SoS = 1420 m/s) (from posterior to anterior). The SoS images were reconstructed based on the different forward models (rows) and four different \textit{a priori} SoS values $\hat{c}$ (columns).}
	\label{Fig:Results_4LayerSoSImage140ms}
\end{figure}

\begin{figure}[!htbp]
    \centering
    \textbf{LVMD phantom}\par\medskip
	\includegraphics[width=1\textwidth]{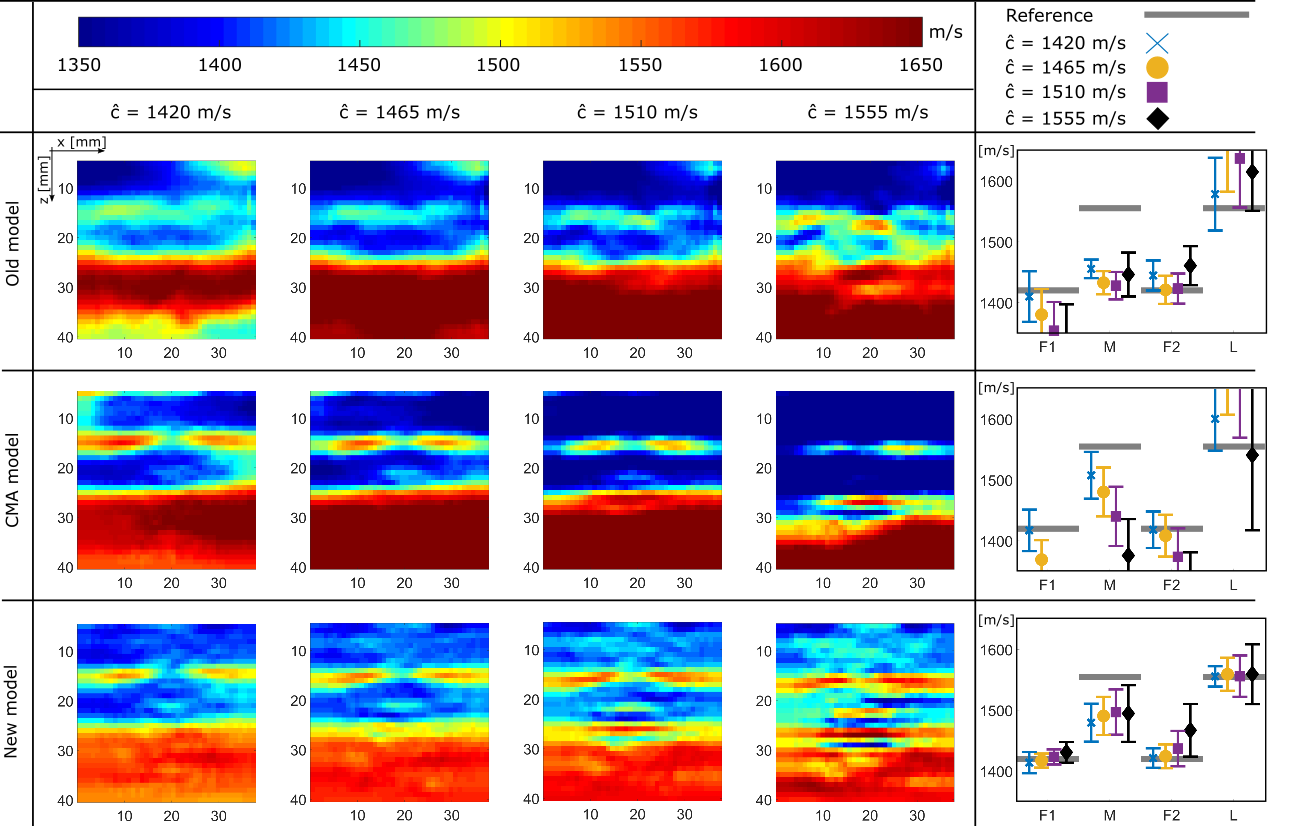}
	\caption{SoS images of the LVMD phantom mimicking the liver (SoS = 1555 m/s) covered by a triangular shaped fat layer (SoS = 1420 m/s), the rectus abdomins (SoS = 1555 m/s) and another thin fat layer (SoS = 1420 m/s) (from posterior to anterior). The SoS images were reconstructed based on the different forward models (rows) and four different \textit{a priori} SoS values $\hat{c}$ (columns).}
	\label{Fig:Results_LiverSoSImage140ms}
\end{figure}

Besides the old and the new model (Eq. \ref{Equ:Materials_OldModel} and \ref{Equ:Materials_NewModel}), an intermediate model was implemented where only the CMA tracking was considered (CMA model, Eq. \ref{Equ:Materials_CMAModel}). The purpose of the CMA model was to assess the influence of the CMA tracking on the results separately from the consideration of the echo position error.\\
The spatial distributions of the reconstructed SoS are showed in color-code (Fig. \ref{Fig:Results_2LayerSoSImage140ms} - \ref{Fig:Results_LiverSoSImage140ms} (rows)). 
The color scale was chosen in a trade-off between covering a large SoS range and allowing to perceive small SoS variations (e.g artefacts). As a consequence, image regions that deviate $\gtrsim$ 70 m/s from the true upper and lower SoS appear saturated.\\
As described in Materials and Methods, the reconstruction of the crf-mode images was based on assuming a uniform SoS value $\hat{c}$. 
To examine the influence of $\hat{c}$ on the final SoS result, crf-mode images were reconstructed with four different $\hat{c}$ covering the range of SoS values between fat-mimicking material (1420 m/s) and muscle/liver tissue mimicking material (1555 m/s) (Fig. \ref{Fig:Results_2LayerSoSImage140ms} - \ref{Fig:Results_LiverSoSImage140ms} columns). The rightmost columns show the mean and standard deviation of the reconstructed SoS within each individual compartment.


\newpage

\newpage

\textbf{Old model}\\
The SoS images reconstructed with the old model (see Fig. \ref{Fig:Results_2LayerSoSImage140ms} to \ref{Fig:Results_LiverSoSImage140ms} - top rows) have in common that they all show artefacts at the top corners, independent of the phantom as well as the \textit{a priori} SoS $\hat{c}$. \\
In the two layer phantom, the SoS of the fat mimicking compartment (F1) strongly depends on the \textit{a priori} SoS $\hat{c}$. Whereas a reasonable SoS - apart form the artefacts - is reconstructed for  $\hat{c}$ of 1420 m/s, higher $\hat{c}$ lead to a distinct underestimation of the SoS in said compartment. In contrast, the mean SoS of the liver mimicking compartment (L) varies only little when changing $\hat{c}$. However, except for $\hat{c}$ of 1465 m/s, the images contain a high level of artefact inside the liver mimicking compartment, suggesting unrealistic SoS inhomogeneities.\\
In the four layer phantom, the mean SoS of the muscle mimicking compartment (M) is reconstructed best for a $\hat{c}$ of 1420 m/s. Higher $\hat{c}$ lead to strong artefacts at the right side of the image, resulting in a decreased mean SoS. Despite these artefacts, however, an increasing $\hat{c}$ leads to an increase in SoS in the middle region of the muscle mimicking compartment. A similar effect can be observed for the post peritoneal fat mimicking compartment (F2), where an increasing $\hat{c}$ leads to a decrease in SoS in the middle region of the layer, but an increase in artefacts at the right side of the image. Inside the liver mimicking compartment, the mean SoS varies between 1541$\pm$34 m/s for a $\hat{c}$ of 1420 m/s and 1571$\pm$40 m/s for a $\hat{c}$ of 1555 m/s. Similar to the two layer phantom, a $\hat{c}$ of 1420 m/s leads to a  decrease in SoS towards the bottom of the image, whereas higher $\hat{c}$ lead to artefacts of high SoS at the bottom of the image. \\
In case of the LVMD phantom, all $\hat{c}$ result in high level of artefact around the wedge-shaped muscle mimicking compartment, making a clear identification difficult. Moreover, regardless of $\hat{c}$, the old model is unable to reconstruct the true SoS of the liver mimicking compartment.\\

\textbf{CMA model}\\
SoS images reconstructed with the CMA model (Fig.  \ref{Fig:Results_2LayerSoSImage140ms} to \ref{Fig:Results_LiverSoSImage140ms} - middle rows) show reasonable mean SoS for the fat mimicking compartments (F1 \& F2) in all phantoms when $\hat{c}$ is close to 1420 m/s. However, independent of the phantom, an increasing $\hat{c}$ leads to a distinct decrease of the reconstructed SoS in any fat mimicking compartment.\\
In the two layer phantom, only a $\hat{c}$ of 1465 m/s leads to reasonable mean SoS in the liver mimicking compartment. Similar to the old model, also in the CMA approach, a $\hat{c}$ of 1420 m/s leads to a distortion of SoS towards the bottom of the image.\\ 
In the four layer phantom, the muscles mimicking compartments mean SoS is overestimated for low $\hat{c}$, whereas high $\hat{c}$ results in an underestimated mean SoS. In contrast, inside the liver mimicking compartment, the mean SoS is underestimated for any $\hat{c}$. Furthermore, similar artefacts toward the bottom of the image as in the two layer phantom are found also in the four layer phantom.\\
Contrary to the old model, the wedge-shaped muscles in the LVMD phantom can be clearly perceived when the reconstruction is performed with the CMA approach. The best agreement of the muscles mimicking compartments SoS with the reference value is found for a $\hat{c}$ of 1420 m/s. An increase in $\hat{c}$, however, leads to a decrease of the reconstructed SoS of the wedge-shaped muscles. Analogous to the old model, the CMA model is, independent of $\hat{c}$, unable to reconstruct the liver mimicking compartments SoS in agreement with the reference value.\\

\textbf{New model}\\
In comparison to the old and the CMA model, the new model leads to substantially improved SoS images among all phantoms, where the best results are achieved when $\hat{c}$ is close to the true SoS of the first fat mimicking layer (F1), i.e. 1420 m/s. For such a $\hat{c}$, the mean SoS of the fat (F1) as well as the liver (L) mimicking compartment deviate only little from the reference value in any phantom. The SoS of the muscle mimicking compartment (M) in the four layer phantom is reconstructed in agreement to the reference value, although only in the middle region of the image. Toward the lateral edges, a decrease in SoS is observed, leading to a decreased mean SoS.  In the LVMD phantom, the mean SoS of the wedge-shaped muscle is underestimated because of the limited spatial resolution of the reconstructed images.\\
Whereas SoS images reconstructed with the old and the CMA model strongly depend on the \textit{a priori} SoS $\hat{c}$, the new model shows a higher stability against variations of the SoS $\hat{c}$. However, also the new model shows a slight increase in the mean SoS when $\hat{c}$ is increased, most pronounced in the top fat mimicking compartment(F1). Furthermore, all images have in common that higher $\hat{c}$ leads to a higher level of artefacts, leading to an increased standard deviation in the individual compartments.

\section{Discussion and Conclusion}
%

Our phantom study revealed two main limitations of the old model in layered samples: First, the absolute SoS values as well as the SoS contrast between different phantom compartments strongly depend on the \textit{a priori} SoS $\hat{c}$. Secondly, in phantoms with lateral SoS variations, i.e. the LVMD phantom, the old CUTE model completely fails in reconstructing the SoS of the different phantom compartments.\\
Two fundamental changes to the CUTE methodology presented in this study lead to the new model that solves these shortcomings. CMA tracking, the first change, determines the echo phase shift not only by a pure Tx-steering approach but in a simultaneous steering of both, the Tx- and the Rx- angle, around a variety of common mid angles. \\
The results showed that when only CMA tracking is implemented, an improvement in the level of artefacts is observed when imaging samples with laterals SoS variations. Thus, whereas in the SoS images determined with the old model the wedge-shaped muscles of the LVMD phantom could not be perceived, they are clearly visible in the SoS images reconstructed with the CMA model. This confirms that there is an implicit Rx-dependence of the aberration delay $\Delta  \tau$ in the phase shift, which was not considered in the old model. However, similar to the old model, also in the CMA model the absolute SoS values as well as the SoS contrast between the different phantom compartments depend strongly on $\hat{c}$. This reveals that the strong dependency of the SoS images on $\hat{c}$ does not origin from a dependency of the receive beamforming on the erroneously assumed background SoS. Anyhow, also the CMA model failed completely in reconstructing the liver's SoS in the LVMD phantom. \\

In addition to the CMA approach, the new model takes into account the phase shift that originates from the erroneous reconstructed position of the echoes. This leads, compared to the other models, to substantially improved quantitative SoS reconstructions among all phantoms. This is most pronounced in the LVMD phantom, where only the new model is able to reconstruct the SoS of the liver compartment close to the reference value, best when $\hat{c}$ is close to the true SoS of the top layer.\\

Even though also the new model shows a slight influence of $\hat{c}$ on the mean SoS of the phantom compartments, it is small compared to the old and the CMA model. We hypothesize that this influence originates from experimental conditions that were not yet accounted for, such as e.g. near-field effects or the 2D approximation of the SoS distribution and sound propagation. Further, an increased level of artefacts in the SoS images is observed when $\hat{c}$ strongly deviates from the SoS of the first layer. This observation is assigned to increased phase noise in the echo phase shift maps, caused by the reduced Tx and Rx focus quality. \\

One important assumption made in this study is the straight ray approximation of US propagation neglecting diffraction and refraction. Given the lateral resolution of the final SoS image (few mm) in relation to the wavelength (0.2 mm at 5 MHz), diffraction will have played a minor role. With the realistically high SoS contrast of the presented phantoms, however, one would expect that refraction cannot be neglected. None-the-less, our results demonstrate that the straight ray approximation was still reasonable in the sense that correct quantitative results could be obtained. Refraction will have a more noteable influence on the reconstructed SoS images in samples with different geometry, e.g spherical structures. Refraction could then be compensated for e.g. using a more accurate forward model, such as an iterative bent-ray approach or the Rytov approximation. \\

Another fundamental assumption is that of sharply defined Tx/Rx angles, but finite angular apertures are needed and were used in the experiments to enable a laterally resolved echo phase shift measurement (and thus SoS image). One may argue that the round-trip time of echoes thus correspond to average slowness integrals over wedge-shaped areas rather than along thin lines. Our results demonstrate that the simplification to line integrals had only minor influence (if any) on the quantitativeness of the SoS. This can be understood from two perspectives: in the first one, the reconstruction of local SoS is based on the local variation of echo phase shift rather than the echo phase shift itself \cite{jaeger2015computed}. Therefore, independent on whether echo phase is influenced by the slowness over a broad or a narrow beam area above the Tx/Rx focus, \textit{at} the focus it is determined by variations of slowness on the spatial scale of the spatial resolution of the focus. From the second perspective, the angular aperture used for Tx/Rx focusing (5$^\circ$)  together with the wavelength (0.2 mm at 5 MHz) result in a focal width of about 3 mm. The active probe aperture needed to focus to the lower edge of the image is about 3.5 mm and thus only slightly larger than the focal width, indicating collimated beams rather than focused ones. At the lateral resolution of echo phase shift detection given by the focal width, the rays can therefore be approximated as lines. \\

\begin{figure}[h]
    \centering
	\includegraphics[width=0.8\textwidth]{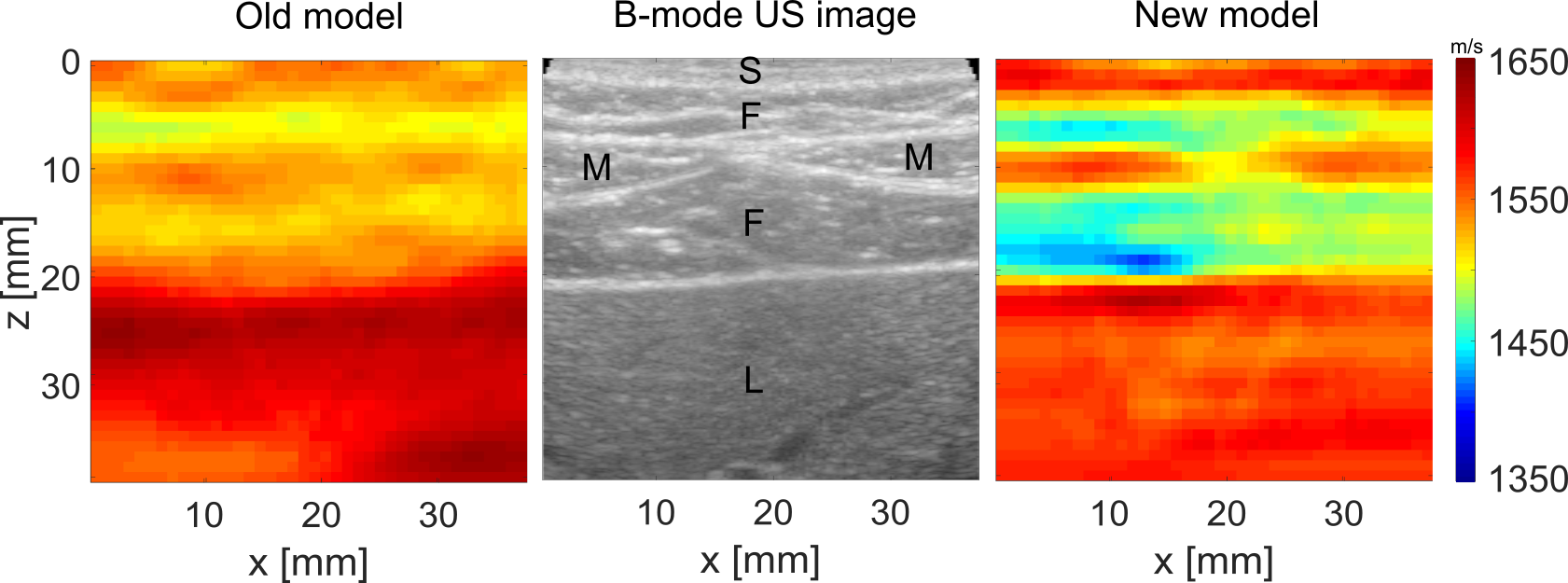}
    \caption{Side-by-side display of conventional B-mode US and SoS image of a healthy volunteer's abdominal wall with subjacent liver tissue, reconstructed with the old and the new model. In the new model, the SoS values of different tissues are well distinguished (S: skin, F: fat, M: rectus abdominis muscle, L: liver parenchyma).}
    \label{Fig:IntroCUTEExample}
\end{figure}

The focus of the present study was to validate the new model. So far we did not attempt to investigate the clinical accuracy of CUTE. We would like to point out, however, that the liver phantoms used in this study were specifically chosen to provide a realistic SoS contrast between different compartments as well as realistic spatial distribution of different compartments. Thus, our liver phantom results suggest that CUTE can achieve sufficient accuracy for a diagnostic application: the new model is able to reconstruct the SoS of the liver mimicking tissue with a deviation from the true SoS of only a few m/s with a standard error of $\approx$ 2.5 m/s in the best and $\approx$ 10 m/s in the worst case. Imbault et al. \cite{imbault2017robust} reported that the average SoS of the liver of healthy patients is $\approx$ 1570 m/s. For patients having steatosis of grade 1 (according to the Brunt scale), the average SoS decreases by about 40 m/s. Thus, CUTE (with the proposed new model) has potential for a quantitative diagnosis of fatty liver disease by complementing real-time b-mode US in a single hand-held device. To illustrate that CUTE can be readily applied to an in-vivo case, Fig. \ref{Fig:IntroCUTEExample} shows the B-mode and SoS images reconstructed with the old and new model of a healthy volunteer's liver. In comparison to the old model, the new model reflects the true anatomical situation and shows a more homogeneous SoS distribution within the liver, as expected in healthy tissue. Moreover, the reconstructed mean SoS of the liver is 1564$\pm$4 m/s and thus reflects healthy liver tissue \cite{lin1987correlations}.\\

Transabdominal hepatic imaging in practise is conventionally performed using curvilinear probes operating at lower frequencies. Whereas this study used a linear probe, the methodology can readily be adapted to curvilinear probes. The advantage of a linear compared to a curvilinear probe is the larger angle range that can be used for Tx/Rx beamsteering, as the same pixel is within the "view" of a smaller number of elements when these elements are pointing into diverging directions. Also, the ratio between the probe aperture length and the image size limits the available angle range in full liver imaging for points that are located far away from the probe. This leads to reduced axial resolution of the SoS image with increasing depth. In situations where one is interested in an average quantitative reading rather than the spatial distribution of SoS, the low axial resolution may not be important. For differential diagnosis of a spatially confined tumor that is located deep inside the liver but well visible on B-mode US, spatial resolution could  be enforced in a prior-guided SoS reconstruction in order to regain quantitative results. In situations where the lack of spatial resolution becomes a problem, using a very large aperture linear array probe instead of a curved one could be a viable solution. \\


In summary, the new model is a key requirement for quantitative handheld reflection-mode SoS imaging. Thanks to the fast angle scanning of US systems and the low computational cost of CUTE, quantitative SoS images can thus be routinely displayed in parallel to conventional B-mode US. CUTE therefore can be used to image any organ that can be examined using echo US, for example the female breast to improve breast cancer diagnostics.  In situations where the echoes are not static, e.g. inside the large blood vessels, blood flow and tissue clutter inhibit phase tracking and may cause SoS artefacts in the reconstruction. To overcome this limitation, we proposed a technique \cite{kuriakose2018receive} where the first order echoes from the moving blood cells are separated from the static echo clutter via a clutter wall filter as in Doppler flow imaging. This technique extends the application of CUTE also to arteries, e.g. for the assessment of plaque inside the carotid artery and may also further improve liver imaging, since it can reduce SoS artefacts which may occur around hepatic veins. Furthermore, knowing the spatial distribution of SoS also allows reconstructing conventional B-mode US images with reduced aberration artefacts \cite{jaeger2015full}.
\section{Acknowledgement}
This research was funded in part by the Swiss National Science Foundation (project number  205320\_178038) and the European Union's Horizon 2020 research and innovation programme under grant agreement No 732411, Photonics Private Public Partnership, and is supported by the Swiss State Secretariat for Education, Research and Innovation (SERI) under contract number 16.0162. The opinions expressed and arguments employed herein do not necessarily reflect the official view of the Swiss Government. The authors thank René Nyffenegger for his technical assistance. 

\newpage

\bibliography{quantitativeSoSImaging} 

\begin{thebibliography}{10}

\bibitem{baker1999sonography}
J.~Baker, P.~J. Kornguth, M.~S. Soo, R.~Walsh, and P.~Mengoni, ``Sonography of
  solid breast lesions: observer variability of lesion description and
  assessment.,'' {\em AJR. American journal of roentgenology}, vol.~172, no.~6,
  pp.~1621--1625, 1999.

\bibitem{konno2001liver}
K.~Konno, H.~Ishida, M.~Sato, T.~Komatsuda, J.~Ishida, H.~Naganuma,
  Y.~Hamashima, and S.~Watanabe, ``Liver tumors in fatty liver: difficulty in
  ultrasonographic interpretation,'' {\em Abdominal imaging}, vol.~26, no.~5,
  pp.~487--491, 2001.

\bibitem{rahbar1999benign}
G.~Rahbar, A.~C. Sie, G.~C. Hansen, J.~S. Prince, M.~L. Melany, H.~E. Reynolds,
  V.~P. Jackson, J.~W. Sayre, and L.~W. Bassett, ``Benign versus malignant
  solid breast masses: Us differentiation,'' {\em Radiology}, vol.~213, no.~3,
  pp.~889--894, 1999.

\bibitem{athanasiou2010breast}
A.~Athanasiou, A.~Tardivon, M.~Tanter, B.~Sigal-Zafrani, J.~Bercoff,
  T.~Deffieux, J.-L. Gennisson, M.~Fink, and S.~Neuenschwander, ``Breast
  lesions: quantitative elastography with supersonic shear
  imaging—preliminary results,'' {\em Radiology}, vol.~256, no.~1,
  pp.~297--303, 2010.

\bibitem{bamber2013efsumb}
J.~Bamber, D.~Cosgrove, C.~Dietrich, J.~Fromageau, J.~Bojunga, F.~Calliada,
  V.~Cantisani, J.-M. Correas, M.~D’onofrio, E.~Drakonaki, {\em et~al.},
  ``Efsumb guidelines and recommendations on the clinical use of ultrasound
  elastography. part 1: Basic principles and technology,'' {\em Ultraschall in
  der Medizin-European Journal of Ultrasound}, vol.~34, no.~02, pp.~169--184,
  2013.

\bibitem{barr2010real}
R.~G. Barr, ``Real-time ultrasound elasticity of the breast: initial clinical
  results,'' {\em Ultrasound quarterly}, vol.~26, no.~2, pp.~61--66, 2010.

\bibitem{cosgrove2013efsumb}
D.~Cosgrove, F.~Piscaglia, J.~Bamber, J.~Bojunga, J.-M. Correas, O.~Gilja,
  A.~Klauser, I.~Sporea, F.~Calliada, V.~Cantisani, {\em et~al.}, ``Efsumb
  guidelines and recommendations on the clinical use of ultrasound
  elastography. part 2: Clinical applications,'' {\em Ultraschall in der
  Medizin-European Journal of Ultrasound}, vol.~34, no.~03, pp.~238--253, 2013.

\bibitem{dietrich2017efsumb}
C.~F. Dietrich, J.~Bamber, A.~Berzigotti, S.~Bota, V.~Cantisani, L.~Castera,
  D.~Cosgrove, G.~Ferraioli, M.~Friedrich-Rust, O.~H. Gilja, {\em et~al.},
  ``Efsumb guidelines and recommendations on the clinical use of liver
  ultrasound elastography, update 2017 (long version),'' {\em Ultraschall in
  der Medizin-European Journal of Ultrasound}, vol.~38, no.~04, pp.~e16--e47,
  2017.

\bibitem{sigrist2017ultrasound}
R.~M. Sigrist, J.~Liau, A.~El~Kaffas, M.~C. Chammas, and J.~K. Willmann,
  ``Ultrasound elastography: review of techniques and clinical applications,''
  {\em Theranostics}, vol.~7, no.~5, p.~1303, 2017.

\bibitem{hu2010photoacoustic}
S.~Hu and L.~V. Wang, ``Photoacoustic imaging and characterization of the
  microvasculature,'' {\em Journal of biomedical optics}, vol.~15, no.~1,
  p.~011101, 2010.

\bibitem{jaeger2012deformation}
M.~Jaeger, D.~C. Harris-Birtill, A.~G. Gertsch-Grover, E.~O’Flynn, and J.~C.
  Bamber, ``Deformation-compensated averaging for clutter reduction in
  epiphotoacoustic imaging in vivo,'' {\em Journal of biomedical Optics},
  vol.~17, no.~6, p.~066007, 2012.

\bibitem{held2016multiple}
K.~G. Held, M.~Jaeger, J.~Ri{\v{c}}ka, M.~Frenz, and H.~G. Akar{\c{c}}ay,
  ``Multiple irradiation sensing of the optical effective attenuation
  coefficient for spectral correction in handheld oa imaging,'' {\em
  Photoacoustics}, vol.~4, no.~2, pp.~70--80, 2016.

\bibitem{Ulrich2018SpectralCorrection}
L.~Ulrich, L.~Ahnen, H.~Akarçay, S.~Sanchez, M.~Jaeger, G.~Held, M.~Wolf, and
  M.~Frenz, ``Spectral correction for handheld optoacoustic imaging by means of
  near‐infrared optical tomography in reflection mode,'' 08 2018.

\bibitem{ruiter2013first}
N.~Ruiter, M.~Zapf, R.~Dapp, T.~Hopp, W.~Kaiser, and H.~Gemmeke, ``First
  results of a clinical study with 3d ultrasound computer tomography,'' in {\em
  2013 IEEE International Ultrasonics Symposium (IUS)}, pp.~651--654, IEEE,
  2013.

\bibitem{greenleaf1981clinical}
J.~F. Greenleaf and R.~C. Bahn, ``Clinical imaging with transmissive ultrasonic
  computerized tomography,'' {\em IEEE Transactions on Biomedical Engineering},
  no.~2, pp.~177--185, 1981.

\bibitem{jago1991experimental}
J.~Jago and T.~Whittingham, ``Experimental studies in transmission ultrasound
  computed tomography,'' {\em Physics in Medicine \& Biology}, vol.~36, no.~11,
  p.~1515, 1991.

\bibitem{zografos2013novel}
G.~Zografos, D.~Koulocheri, P.~Liakou, M.~Sofras, S.~Hadjiagapis, M.~Orme, and
  V.~Marmarelis, ``Novel technology of multimodal ultrasound tomography detects
  breast lesions,'' {\em European radiology}, vol.~23, no.~3, pp.~673--683,
  2013.

\bibitem{carson1981breast}
P.~L. Carson, C.~R. Meyer, A.~L. Scherzinger, and T.~V. Oughton, ``Breast
  imaging in coronal planes with simultaneous pulse echo and transmission
  ultrasound,'' {\em Science}, vol.~214, no.~4525, pp.~1141--1143, 1981.

\bibitem{huthwaite2011high}
P.~Huthwaite and F.~Simonetti, ``High-resolution imaging without iteration: A
  fast and robust method for breast ultrasound tomography,'' {\em The Journal
  of the Acoustical Society of America}, vol.~130, no.~3, pp.~1721--1734, 2011.

\bibitem{wiskin2012non}
J.~Wiskin, D.~Borup, S.~Johnson, and M.~Berggren, ``Non-linear inverse
  scattering: High resolution quantitative breast tissue tomography,'' {\em The
  Journal of the Acoustical Society of America}, vol.~131, no.~5,
  pp.~3802--3813, 2012.

\bibitem{sandhu2015frequency}
G.~Sandhu, C.~Li, O.~Roy, S.~Schmidt, and N.~Duric, ``Frequency domain
  ultrasound waveform tomography: breast imaging using a ring transducer,''
  {\em Physics in Medicine \& Biology}, vol.~60, no.~14, p.~5381, 2015.

\bibitem{hesse2013nonlinear}
M.~C. Hesse, L.~Salehi, and G.~Schmitz, ``Nonlinear simultaneous reconstruction
  of inhomogeneous compressibility and mass density distributions in
  unidirectional pulse-echo ultrasound imaging,'' {\em Physics in Medicine and
  Biology}, vol.~58, no.~17, p.~6163, 2013.

\bibitem{shin2010estimation}
H.-C. Shin, R.~Prager, H.~Gomersall, N.~Kingsbury, G.~Treece, and A.~Gee,
  ``Estimation of average speed of sound using deconvolution of medical
  ultrasound data,'' {\em Ultrasound in medicine \& biology}, vol.~36, no.~4,
  pp.~623--636, 2010.

\bibitem{krucker2004sound}
J.~Krucker, J.~B. Fowlkes, and P.~L. Carson, ``Sound speed estimation using
  automatic ultrasound image registration,'' {\em IEEE transactions on
  ultrasonics, ferroelectrics, and frequency control}, vol.~51, no.~9,
  pp.~1095--1106, 2004.

\bibitem{imbault2017robust}
M.~Imbault, A.~Faccinetto, B.-F. Osmanski, A.~Tissier, T.~Deffieux, J.-L.
  Gennisson, V.~Vilgrain, and M.~Tanter, ``Robust sound speed estimation for
  ultrasound-based hepatic steatosis assessment,'' {\em Physics in Medicine and
  Biology}, vol.~62, no.~9, p.~3582, 2017.

\bibitem{kondo1990evaluation}
M.~Kondo, K.~Takamizawa, M.~Hirama, K.~Okazaki, K.~Iinuma, and Y.~Takehara,
  ``An evaluation of an in vivo local sound speed estimation technique by the
  crossed beam method,'' {\em Ultrasound in Medicine and Biology}, vol.~16,
  no.~1, pp.~65--72, 1990.

\bibitem{cespedes1992feasibility}
I.~C{\'e}spedes, J.~Ophir, and Y.~Huang, ``On the feasibility of pulse-echo
  speed of sound estimation in small regions: Simulation studies,'' {\em
  Ultrasound in Medicine and Biology}, vol.~18, no.~3, pp.~283--291, 1992.

\bibitem{jakovljevic2018local}
M.~Jakovljevic, S.~Hsieh, R.~Ali, G.~Chau Loo~Kung, D.~Hyun, and J.~J. Dahl,
  ``Local speed of sound estimation in tissue using pulse-echo ultrasound:
  Model-based approach,'' {\em The Journal of the Acoustical Society of
  America}, vol.~144, no.~1, pp.~254--266, 2018.

\bibitem{jaeger2015computed}
M.~Jaeger, G.~Held, S.~Peeters, S.~Preisser, M.~Gr{\"u}nig, and M.~Frenz,
  ``Computed ultrasound tomography in echo mode for imaging speed of sound
  using pulse-echo sonography: proof of principle,'' {\em Ultrasound in
  medicine and biology}, vol.~41, no.~1, pp.~235--250, 2015.

\bibitem{jaeger2015towards}
M.~Jaeger and M.~Frenz, ``Towards clinical computed ultrasound tomography in
  echo-mode: Dynamic range artefact reduction,'' {\em Ultrasonics}, vol.~62,
  pp.~299--304, 2015.

\bibitem{loupas1995axial}
T.~Loupas, J.~Powers, and R.~W. Gill, ``An axial velocity estimator for
  ultrasound blood flow imaging, based on a full evaluation of the doppler
  equation by means of a two-dimensional autocorrelation approach,'' {\em IEEE
  transactions on ultrasonics, ferroelectrics, and frequency control}, vol.~42,
  no.~4, pp.~672--688, 1995.

\bibitem{flax1988phase}
S.~Flax and M.~O'Donnell, ``Phase-aberration correction using signals from
  point reflectors and diffuse scatterers: Basic principles,'' {\em IEEE
  transactions on ultrasonics, ferroelectrics, and frequency control}, vol.~35,
  no.~6, pp.~758--767, 1988.

\bibitem{nock1989phase}
L.~Nock, G.~E. Trahey, and S.~W. Smith, ``Phase aberration correction in
  medical ultrasound using speckle brightness as a quality factor,'' {\em The
  Journal of the Acoustical Society of America}, vol.~85, no.~5,
  pp.~1819--1833, 1989.

\bibitem{rachlin1990direct}
D.~Rachlin, ``Direct estimation of aberrating delays in pulse-echo imaging
  systems,'' {\em The Journal of the Acoustical Society of America}, vol.~88,
  no.~1, pp.~191--198, 1990.

\bibitem{li1997phaseA}
Y.~Li, ``Phase aberration correction using near-field signal redundancy. i.
  principles [ultrasound medical imaging],'' {\em IEEE transactions on
  ultrasonics, ferroelectrics, and frequency control}, vol.~44, no.~2,
  pp.~355--371, 1997.

\bibitem{li1997phaseB}
Y.~Li, D.~Robinson, and D.~Carpenter, ``Phase aberration correction using
  near-field signal redundancy. ii. experimental results,'' {\em ieee
  transactions on ultrasonics, ferroelectrics, and frequency control}, vol.~44,
  no.~2, pp.~372--379, 1997.

\bibitem{haun2004overdetermined}
M.~A. Haun, D.~L. Jones, and W.~Oz, ``Overdetermined least-squares aberration
  estimates using common-midpoint signals,'' {\em IEEE transactions on medical
  imaging}, vol.~23, no.~10, pp.~1205--1220, 2004.

\bibitem{jaeger2015full}
M.~Jaeger, E.~Robinson, H.~G. Akar{\c{c}}ay, and M.~Frenz, ``Full correction
  for spatially distributed speed-of-sound in echo ultrasound based on
  measuring aberration delays via transmit beam steering,'' {\em Physics in
  medicine \& biology}, vol.~60, no.~11, p.~4497, 2015.

\bibitem{MJTalk2015}
M.~Jaeger and M.~Frenz, ``Quantitative imaging of speed of sound in echo
  ultrasonography.'' IEEE International Ultrasound Symposium, Taipei, 2015.

\bibitem{sanabria2018spatial}
S.~J. Sanabria, E.~Ozkan, M.~Rominger, and O.~Goksel, ``Spatial domain
  reconstruction for imaging speed-of-sound with pulse-echo ultrasound:
  simulation and in vivo study,'' {\em Physics in Medicine \& Biology},
  vol.~63, no.~21, p.~215015, 2018.

\bibitem{bednar2009modeling}
J.~B. Bednar, ``Modeling, migration and velocity analysis in simple and complex
  structure,'' {\em by Panorama Technologies, Inc}, 2009.

\bibitem{bercoff2011ultrafast}
J.~Bercoff, ``Ultrafast ultrasound imaging,'' in {\em Ultrasound
  imaging-Medical applications}, IntechOpen, 2011.

\bibitem{montaldo2009coherent}
G.~Montaldo, M.~Tanter, J.~Bercoff, N.~Benech, and M.~Fink, ``Coherent
  plane-wave compounding for very high frame rate ultrasonography and transient
  elastography,'' {\em IEEE transactions on ultrasonics, ferroelectrics, and
  frequency control}, vol.~56, no.~3, pp.~489--506, 2009.

\bibitem{madsen1982anthropomorphic}
E.~L. Madsen, J.~A. Zagzebski, and G.~R. Frank, ``An anthropomorphic ultrasound
  breast phantom containing intermediate-sized scatteres,'' {\em Ultrasound in
  Medicine and Biology}, vol.~8, no.~4, pp.~381--392, 1982.

\bibitem{madsen2003tissue}
E.~Madsen, G.~Frank, T.~Krouskop, T.~Varghese, F.~Kallel, and J.~Ophir,
  ``Tissue-mimicking oil-in-gelatin dispersions for use in heterogeneous
  elastography phantoms,'' {\em Ultrasonic imaging}, vol.~25, no.~1,
  pp.~17--38, 2003.

\bibitem{madsen2006anthropomorphic}
E.~L. Madsen, M.~A. Hobson, G.~R. Frank, H.~Shi, J.~Jiang, T.~J. Hall,
  T.~Varghese, M.~M. Doyley, and J.~B. Weaver, ``Anthropomorphic breast
  phantoms for testing elastography systems,'' {\em Ultrasound in medicine and
  biology}, vol.~32, no.~6, pp.~857--874, 2006.

\bibitem{nguyen2014development}
M.~M. Nguyen, S.~Zhou, J.-l. Robert, V.~Shamdasani, and H.~Xie, ``Development
  of oil-in-gelatin phantoms for viscoelasticity measurement in ultrasound
  shear wave elastography,'' {\em Ultrasound in medicine and biology}, vol.~40,
  no.~1, pp.~168--176, 2014.

\bibitem{lin1987correlations}
T.~Lin, J.~Ophir, and G.~Potter, ``Correlations of sound speed with tissue
  constituents in normal and diffuse liver disease,'' {\em Ultrasonic imaging},
  vol.~9, no.~1, pp.~29--40, 1987.

\bibitem{kuriakose2018receive}
M.~Kuriakose, J.-W. Muller, P.~St{\"a}hli, M.~Frenz, and M.~Jaeger, ``Receive
  beam-steering and clutter reduction for imaging the speed-of-sound inside the
  carotid artery,'' {\em Journal of Imaging}, vol.~4, no.~12, p.~145, 2018.

\end{thebibliography}

\bibliographystyle{ieeetr}

\end{document}